\documentclass[10pt]{article}
\usepackage{multicol}
\usepackage{graphicx}
\usepackage{amsmath}
\usepackage[a4paper]{geometry}
\usepackage{hyperref}
\usepackage{rotating}

\renewcommand{\baselinestretch}{1.15}

\begin{document}

\begin{center}
{\Large \bf Excitation functions of related parameters from
transverse momentum (mass) spectra in high energy collisions}

\vskip.75cm

Li-Li~Li$^1$, Fu-Hu~Liu$^{1,}${\footnote{E-mail: fuhuliu@163.com;
fuhuliu@sxu.edu.cn}}, Muhammad~Waqas$^1$, Rasha~Al-Yusufi$^{1,2}$,
Altaf~Mujear$^{1,2}$

\vskip.25cm

{\small\it $^1$Institute of Theoretical Physics and State Key
Laboratory of Quantum Optics and Quantum Optics Devices,\\ Shanxi
University, Taiyuan, Shanxi 030006, China

$^2$Physics Department, Faculty of Science, Sana'a University, P.
O. Box 13499, Sana'a, Republic of Yemen}

\end{center}

\vskip.5cm

{\bf Abstract:} Transverse momentum (mass) spectra of positively
and negatively charged pions, positively and negatively charged
kaons, protons and antiprotons produced at mid-(pseudo)rapidity in
various collisions at high energies are analyzed in this work. The
experimental data measured in central gold-gold, central
lead-lead, and inelastic proton-proton collisions by several
international collaborations are studied. The (two-component)
standard distribution is used to fit the data and extract the
excitation function of effective temperature. Then, the excitation
functions of kinetic freeze-out temperature, transverse flow
velocity, and initial temperature are obtained. In the considered
collisions, the four parameters increase with the increase of
collision energy in general, and the kinetic freeze-out
temperature appears the trend of saturation at the top
Relativistic Heavy Ion Collider and the Large Hadron Collider.
\\

\\


\vskip1.0cm

\begin{multicols}{2}

{\section{Introduction}}

It is believed that the environment of high temperature and high
density is formed in the system evolution process of central
nucleus-nucleus ($AA$) collisions at high energy~\cite{1,2,3}, in
which quark-gluon plasma (QGP) is possibly created and many
particles are produced~\cite{4,5,6}. At present, it is impossible
to detect directly the system evolution process of collisions due
to very short time interval. Instead, the particle spectra at the
stage of kinetic freeze-out can be measured in experiments and the
mechanisms of system evolutions and particle productions can be
studied indirectly~\cite{7,8,9}, though the particle ratios
reflect the property at the stage of chemical freeze-out. As for
peripheral $AA$ collisions and small collision system, the
situation is similar if the multiplicity is high enough due to
small system also appears collective behavior~\cite{10,11}.

Although there are different stages in the system
evolution~\cite{1,2,3}, the initial state is the most important
due to its determining effect to the system evolution. In
addition, chemical and kinetic freeze-outs are two important
stages in the system evolution. At the stage of chemical
freeze-out, the system had happened the phase transition from QGP
to hadronic matter, and the constituents and ratios of various
particles do not change anymore. At the stage of kinetic
freeze-out, the collisions among various particles are elastic,
and the transverse momentum spectra of various particles are
fixed~\cite{2,7}. In small system with low multiplicity, QGP is
not expected to create in it due to a very small volume of the
violent collision region. From the similar multiplicity at the
energy up to 200 GeV, small system is more similar to peripheral
$AA$ collisions, but not to central $AA$ collisions~\cite{12,12a}.
At the energy down to 10 or several GeV, the situation is
different due to the fact that baryon-dominated effect plays more
important role in $AA$ collisions~\cite{12b}.

The temperatures at the stages of kinetic freeze-out, chemical
freeze-out, and initial state are called the kinetic freeze-out
temperature ($T_0$ or $T_{kin}$), chemical freeze-out temperature
($T_{ch}$), and initial temperature ($T_i$), respectively.
Besides, one also has the effective temperature ($T$) in which
both the contributions of thermal motion and flow effect are
included. It is expected that various temperatures can be
extracted from particle spectra, which are usually model
dependent. Generally, $T$ is unavoidably model dependent, and
$T_{ch}$ extracted from particle ratios in the statistical thermal
model~\cite{13,14,15,16} is also model dependent. We hope to use a
less model dependent method to extract $T_0$, $\beta_T$, and
$T_i$. The quantities used in the method are expected to relate to
experimental data as much as possible, though they can be
calculated from models in some cases.

To perform a less model dependent method, we would like to use the
standard distribution or its two-component form to obtain $T$ by
fitting the experimental transverse momentum ($p_T$) or transverse
mass ($m_T$) spectra of various particles. The standard
distribution includes the Bose-Einstein, Fermi-Dirac, and
Boltzmann distributions, in which the effective temperature
parameter $T$ are the closest to that in the ideal gas model when
comparing $T$ with those in other distributions. After the
fitting, we hope to extract $T_0$ and $\beta_T$ from the relation
to average $p_T$ ($\langle p_T\rangle$) due to the Erlang
distribution in multisource thermal model~\cite{17,18,19} and
$T_i$ from the relation to root-mean-square $p_T$ ($\sqrt{\langle
p_T^2\rangle}$) due to the color string percolation
model~\cite{20,21,22}. Obviously, $\langle p_T\rangle$ and
$\sqrt{\langle p_T^2\rangle}$ depend on the data themselves,
though they can be calculated from models.

In this work, the $p_T$ ($m_T$) spectra of positively and
negatively charged pions ($\pi^+$ and $\pi^-$), positively and
negatively charged kaons ($K^+$ and $K^-$), protons and
antiprotons ($p$ and $\bar p$) produced at mid-(pseudo)rapidity
(mid-$y$ or mid-$\eta$) measured in central gold-gold (Au-Au)
collisions at the Alternating Gradient Synchrotron (AGS) by the
E866~\cite{23}, E895~\cite{24,25}, and E802~\cite{26,27}
Collaborations and at the Relativistic Heavy Ion Collider (RHIC)
by the STAR~\cite{28,29,30} and PHENIX~\cite{31,32}
Collaborations, in central lead-lead (Pb-Pb) collisions at the
Super Proton Synchrotron (SPS) by the NA49
Collaboration~\cite{33,34,35} and at the Large Hadron Collider
(LHC) by the ALICE Collaboration~\cite{36}, as well as in
inelastic (INEL) proton-proton ($pp$) collisions at the SPS by the
NA61/SHINE Collaboration~\cite{37,38}, at the RHIC by the PHENIX
Collaboration~\cite{39}, and at the LHC by the CMS
Collaboration~\cite{40,41} are studied. The (two-component)
standard distribution is used to fit the data and to extract $T$,
$T_i$, $T_0$, and $\beta_T$, as well as the excitation functions
(energy dependences) of parameters.

The remainder of this paper is structured as follows. The
formalism and method are shortly described in Section 2. Results
and discussion are given in Section 3. In Section 4, we summarize
our main observations and conclusions.
\\

{\section{Formalism and method}}

In high energy collisions, the soft excitation and hard scattering
processes are two main processes of particle productions. Most
light flavor particles are produced in the soft excitation process
and distribute in a narrow $p_T$ range which is less than $2\sim3$
GeV/$c$ or a little more. Some light flavor particles are produced
in the hard scattering process and distribute in a wide $p_T$
range. In collisions at not too high energies, the contribution of
hard scattering process can be neglected and the main contributor
to produced particles is the soft excitation process. In
collisions at high energy, the contribution of hard scattering
process cannot be neglected, though the main contributor to
produced particles is also the soft excitation process. It is
expected that the contribution fraction of hard scattering process
increases with the increase of collisions energy.

The contributions of soft excitation and hard scattering processes
can be described by similar or different probability density
functions. Generally, the hard scattering process does not
contribute mainly to the temperature and flow velocity due to its
small fraction in a narrow $p_T$ range. We can neglect the
contribution of hard scattering process if we study the spectra in
a not too wide $p_T$ range. On the contribution of soft excitation
process, we have more than one functions to describe the $p_T$
spectra. These functions include, but are not limited to, the
standard distribution~\cite{42}, the Tsallis
statistics~\cite{42,43,44,45}, the Erlang
distribution~\cite{17,18,19}, the Schwinger
mechanism~\cite{46,47,48,49}, the blast-wave model with Boltzmann
statistics~\cite{50,51}, the blast-wave model with Tsallis
statistics~\cite{52,53,54}, the Hagedorn thermal
distribution~\cite{55}, and their superposition with two- or
three-component. These functions also describe partly the $p_T$
spectra of hard scattering process in most cases.

In our opinion, in the case of fitting the data with acceptable
representations, various distributions show similar behaviors
which result in similar $\langle p_T\rangle$ ($\sqrt{\langle
p_T^2\rangle}$) with different parameters. To be the closest to
the temperature concept in the ideal gas model, we choose the
standard distribution in which the chemical potential $\mu$ and
spin property $S$ are included. That is, one has the probability
density function in terms of $p_T$ to be~\cite{42}
\begin{align}
f_{p_T}(p_T,T) &=\frac{1}{N}\frac{dN}{dp_T}= C p_T m_T
\int_{y_{\min}}^{y_{\max}} \cosh y \nonumber\\
& \times \left[\exp\left(\frac{m_T\cosh y-\mu}{T} \right)+S
\right]^{-1} dy,
\end{align}
where
\begin{align}
m_T=\sqrt{p_T^2+m_0^2},
\end{align}
$m_0$ is the rest mass, $N$ denotes the particle number,
$y_{\min}$ ($y_{\max}$) is the minimum (maximum) value in the
rapidity interval, $S=-1$ (+1) is for bosons (fermions), and $C$
is the normalization constant. Similarly, the probability density
function in terms of $m_T$ is
\begin{align}
f_{m_T}(m_T,T) &=\frac{1}{N}\frac{dN}{dm_T}= C m_T^2
\int_{y_{\min}}^{y_{\max}} \cosh y \nonumber\\
& \times \left[\exp\left(\frac{m_T\cosh y-\mu}{T} \right)+S
\right]^{-1} dy.
\end{align}
In some cases, the independent variable $m_T$ in Eq. (3) is
replaced by $m_T-m_0$ which starts at 0. Both $m_T$ and $m_T-m_0$
show the same distribution shape. As probability density
functions, the integrals of Eqs. (1) and (3) are naturally
normalized to 1 respectively.

The chemical potential $\mu$ in Eqs. (1) and (3) is particle
dependent. For the particle type $i$ ($i=\pi$, $K$, and $p$ in
this work), its chemical potential $\mu_i$ is expressed
by~\cite{32,56,57}
\begin{align}
\mu_i=-\frac{1}{2}T_{ch} \ln\left( k_i \right),
\end{align}
where $k_i$ denotes the ratio of negative to positive particle
numbers,
\begin{align}
T_{ch}=\frac{T_{\lim}}{1+\exp\left[2.60- \ln \left(
\sqrt{s_{NN}}\right) /0.45\right]}
\end{align}
is empirically the chemical freeze-out temperature in the
statistical thermal model~\cite{13,14,15,16}, $T_{\lim}=0.158$ GeV
is the limiting or saturation temperature~\cite{3}, and
$\sqrt{s_{NN}}$ is the center-of-mass energy per nucleon pair in
the units of GeV.

Generally, one needs one or two standard distributions to fit the
$p_T$ ($m_T$) spectra in a narrow range. In particular, if the
resonance decays contribute a large fraction, a two-component
distribution is indeed needed. Or, if the hard scattering process
contributes a sizable fraction in the considered $p_T$ ($m_T$)
range, a two-component distribution is also needed. In the case of
using the two-component standard distribution in which the
contributions from resonance decay are naturally included in the
first component which covers the spectra in low-$p_T$ region
($<0.2\sim0.3$ GeV/$c$), one has the probability density functions
of $p_T$ and $m_T$ to be
\begin{align}
f_{p_T}(p_T)=kf_{p_T}(p_T,T_1)+(1-k)f_{p_T}(p_T,T_2)
\end{align}
and
\begin{align}
f_{m_T}(m_T)=kf_{m_T}(m_T,T_1)+(1-k)f_{m_T}(m_T,T_2)
\end{align}
respectively, where $k$ denotes the contribution fraction of the
first component, and $f_{p_T}(p_T,T_1)$ [$f_{p_T}(p_T,T_2)$] and
$f_{m_T}(m_T,T_1)$ [$f_{m_T}(m_T,T_2)$] are given in Eqs. (1) and
(3) respectively. The integrals of Eqs. (6) and (7) are also
normalized to 1 respectively. Correspondingly,
\begin{align}
T=kT_1+(1-k)T_2
\end{align}
is averaged by weighting the two fractions. The temperature $T$
defined in Eq. (8) reflects the common effective temperature of
the two components in the case of the two components are assumed
to stay in equilibrium.

According to the Hagedorn model~\cite{55}, one may also use the
usual step function $\theta(x)$ to superpose the two standard
distributions, where $\theta(x)=0$ if $x<0$ and $\theta(x)=1$ if
$x\geq0$. Thus, we have new probability density functions of $p_T$
and $m_T$ to be
\begin{align}
f_{p_T}(p_T) &=A_1\theta(p_1-p_T)f_{p_T}(p_T,T_1) \nonumber\\
&+A_2\theta(p_T-p_1) f_{p_T}(p_T,T_2)
\end{align}
and
\begin{align}
f_{m_T}(m_T) &=A_1\theta(m_1-m_T)f_{m_T}(m_T,T_1) \nonumber\\
&+A_2\theta(m_T-m_1) f_{m_T}(m_T,T_2)
\end{align}
respectively, where $A_1$ and $A_2$ are constants which result in
the two components to be equal to each other at $p_T=p_1$ and
$m_T=m_1$. The integrals of Eqs. (9) and (10) should be normalized
to 1 respectively due to the fact that they are probability
density functions. The contribution fractions of the first
component in Eqs. (9) and (10) are
\begin{align}
k &=\int_0^{p_1} \hskip-3mm A_1f_{p_T}(p_T,T_1)dp_T \nonumber\\
&=1- \int_{p_1}^{p_{T\max}} \hskip-3mm A_2f_{p_T}(p_T,T_2)dp_T
\end{align}
and
\begin{align}
k &=\int_{m_0}^{m_1} \hskip-3mm A_1f_{m_T}(m_T,T_1)dm_T \nonumber\\
&=1- \int_{m_1}^{m_{T\max}} \hskip-3mm A_2f_{m_T}(m_T,T_2)dm_T
\end{align}
respectively, where $p_{T\max}$ and $m_{T\max}$ denote the maximum
$p_T$ and $m_T$ respectively. Eq. (8) is also suitable for the
superposition in terms of the Hagedorn model~\cite{55}.

The two superpositions show respective advantages and
disadvantages. The first superposition can fit the data by a
smooth curve. However, there are correlations in determining $T_1$
and $T_2$. The second superposition can determine $T_1$ and $T_2$
without correlations. However, the curves are possibly not smooth
at $p_1$ or $m_1$. In the case of obtaining $\langle p_T\rangle$
and $\sqrt{\langle p_T^2\rangle}$, it does not matter which
superposition is used, though the two $T$ are slightly different.
In this work, we use the first superposition to obtain smooth
curves. One has
\begin{align}
\langle p_T\rangle =\int_0^{p_{T\max}}p_Tf_{p_T}(p_T)dp_T
\end{align}
and
\begin{align}
\sqrt{\langle p_T^2\rangle} =\sqrt{\int_0^{p_{T\max}}p_T^2
f_{p_T}(p_T)dp_T}
\end{align}
due to
\begin{align}
\int_0^{p_{T\max}} \hskip-3mm f_{p_T}(p_T)dp_T = 1.
\end{align}
Based on $m_T$ spectrum, we may use the same parameters to obtain
$\langle p_T\rangle$ and $\sqrt{\langle p_T^2\rangle}$ from the
related formula of $p_T$ distribution.

It should be noted that, since we aim to extract the parameters in
a less model dependent way, we shall obtain $\langle p_T\rangle$
and $\sqrt{\langle p_T^2\rangle}$ from the combination of data
points and fit function in this paper. In fact, we may divide
$p_T$ ($m_T$) spectrum into two or three regions according to the
measured and un-measured $p_T$ ($m_T$) ranges. To obtain $\langle
p_T\rangle$ and $\sqrt{\langle p_T^2\rangle}$, we may use the data
points in the measured $p_T$ ($m_T$) range and only use the fit
function to extrapolate to the un-measured $p_T$ ($m_T$) range.

In each nucleon-nucleon collision in $AA$ and $pp$ collisions, the
projectile and target participant sources contribute equally to
$\langle p_T\rangle$. In the framework of multisource thermal
model~\cite{17,18,19}, each projectile and target source
contribute a fraction of 1/2 to $\langle p_T\rangle$, i.e.
$\langle p_T\rangle/2$ which is contributed together by the
thermal motion and flow effect. Let $k_0$ ($1-k_0$) denote the
contribution fraction of thermal motion (flow effect), we define
empirically
\begin{align}
T_0\equiv \frac{k_0\langle p_T\rangle}{2}
\end{align}
and
\begin{align}
\beta_T\equiv \frac{(1-k_0)\langle p_T\rangle}{2m_0 \overline
\gamma},
\end{align}
where $\overline\gamma$ is the mean Lorentz factor of the
considered particles and
\begin{align}
k_0 = 0.30-0.01\ln\left(\sqrt{s_{NN}}\right)
\end{align}
is a parameterized representation in this paper due to our
comparison with the results~\cite{12,12a} from the blast-wave
model~\cite{50,51,52,53,54}. In Eq. (18), $\sqrt{s_{NN}}$ is in
the units of GeV as that in Eq. (5).

In a recent work~\cite{57a}, it is shown that the effective
temperature is proportional to $\langle p_T\rangle$ and the
kinetic freeze-out temperature is proportional to the effective
temperature, though the effective temperature used in
ref.~\cite{57a} is different from this paper. This confirms the
relation of $T_0\propto \langle p_T\rangle$ (Eq. (16)) used in
this paper. Considering each projectile and target source
contributing $\langle p_T\rangle/2$~\cite{17,18,19}, we have
concretely $T_0\propto \langle p_T\rangle/2$. The remainder in
$\langle p_T\rangle/2$ is naturally contributed by transverse
flow. This confirms Eqs. (16) and (17) to be justified, though
$k_0$ is an empirical representation.

To continue this work, we need some assumptions and a coordinate
system. In the source rest frame, the particles are assumed to
emit isotropically. Meanwhile, the interactions among various
sources are neglected, which affects slightly the $p_T$ ($m_T$)
spectra, though which affects largely anisotropic flows~\cite{58}.
A right-handed coordinate system $O$--$xyz$ is established in the
source rest frame, where $Oz$ axis is along the beam direction,
$xOy$ plane is the transverse plane, and $xOz$ plane is the
reaction plane.

We can obtain $\overline\gamma$ by a Monte Carlo (MC) method. Let
$R_{1,2,3}$ denote random number distributed evenly in $[0,1]$,
each concrete $p_T$ satisfies
\begin{align}
\int_0^{p_T}\hskip-3mm f_{p_T}(p'_T,T)dp'_T <R_1<
\int_0^{p_T+\delta p_T}\hskip-3mm f_{p_T}(p'_T,T)dp'_T,
\end{align}
where $\delta p_T$ denotes a small shift relative to $p_T$. Each
concrete emission angle $\theta$ satisfies
\begin{align}
\theta=2\arcsin\sqrt{R_2}
\end{align}
due to the fact that $\theta$ obeys the probability density
function $f_{\theta}(\theta)=(1/2)\sin\theta$ in $[0,\pi]$ in the
case of isotropic assumption in the source rest frame. The
solution of the equation
$\int_0^{\theta}f_{\theta'}(\theta')d\theta'=R_2$ is Eq. (20). We
give up to use rapidity due to the fact that it is unnecessary
here. Each concrete momentum $p$, energy $E$, and Lorentz factor
$\gamma$ can be obtained by
\begin{align}
p=p_T \csc\theta,
\end{align}
\begin{align}
E=\sqrt{p^2+m_0^2}
\end{align}
and
\begin{align}
\gamma=\frac{E}{m_0},
\end{align}
respectively. After multiple repeating calculations due to the MC
method, we have
\begin{align}
\overline\gamma=\frac{\overline E}{m_0},
\end{align}
where $\overline E$ denotes the mean $E$ for a given type of
particle.

In addition, each concrete azimuthal angle $\phi$ satisfies
\begin{align}
\phi=2\pi R_3
\end{align}
due to the fact that $\phi$ obeys the probability density function
$f_{\phi}(\phi)=1/(2\pi)$ in $[0,2\pi]$ in the case of isotropic
assumption in the source rest frame. The solution of the equation
$\int_0^{\phi}f_{\phi'}(\phi')d\phi'=R_3$ is Eq. (25). Each
concrete momentum components $p_x$, $p_y$, and $p_z$ can be
obtained by
\begin{align}
p_x=p_T \cos\phi,
\end{align}
\begin{align}
p_y=p_T \sin\phi,
\end{align}
and
\begin{align}
p_z=p_T \cot\theta =p\cos\theta,
\end{align}
respectively. By using the components and $E$, $p$, and $\theta$,
we can obtain other quantities such as (pseudo)rapidity and event
structure~\cite{58} which are beyond the focus of this work and
will not be studied anymore.

According to the color string percolation model~\cite{20,21,22},
one has
\begin{align}
T_i\equiv \sqrt{\frac{\langle p_T^2\rangle}{2}}.
\end{align}
Meanwhile, we have the relation between the three components
$p_x$, $p_y$, and $p_z$ of the momentum $p$ to be
\begin{align}
\sqrt{\langle p_x^2\rangle} =\sqrt{\langle p_y^2\rangle}
=\sqrt{\langle p_z^2\rangle} = \sqrt{\frac{\langle
p_T^2\rangle}{2}},
\end{align}
in which the root-mean-square components $\sqrt{\langle
p_x^2\rangle}$, $\sqrt{\langle p_y^2\rangle}$, and $\sqrt{\langle
p_z^2\rangle}$ are used. Naturally, $T_i$ can be given by one of
the root-mean-square components.

We would like to point out that the above isotropic assumption is
only performed in the source rest frame. It is expected that many
sources are formed in high energy collisions according to the
multisource model~\cite{17,18,19}. These sources distribute at
different rapidities in the rapidity space, which appear the
effect of longitudinal flow. The two-component $p_T$ and $m_T$
spectra render that these sources stay in two different excitation
states or have two different decay mechanisms. The interactions
among these sources also affect anisotropic flows in transverse
plane~\cite{58}.
\\

\begin{figure*}[!htb]
\begin{center}
\includegraphics[width=16.0cm]{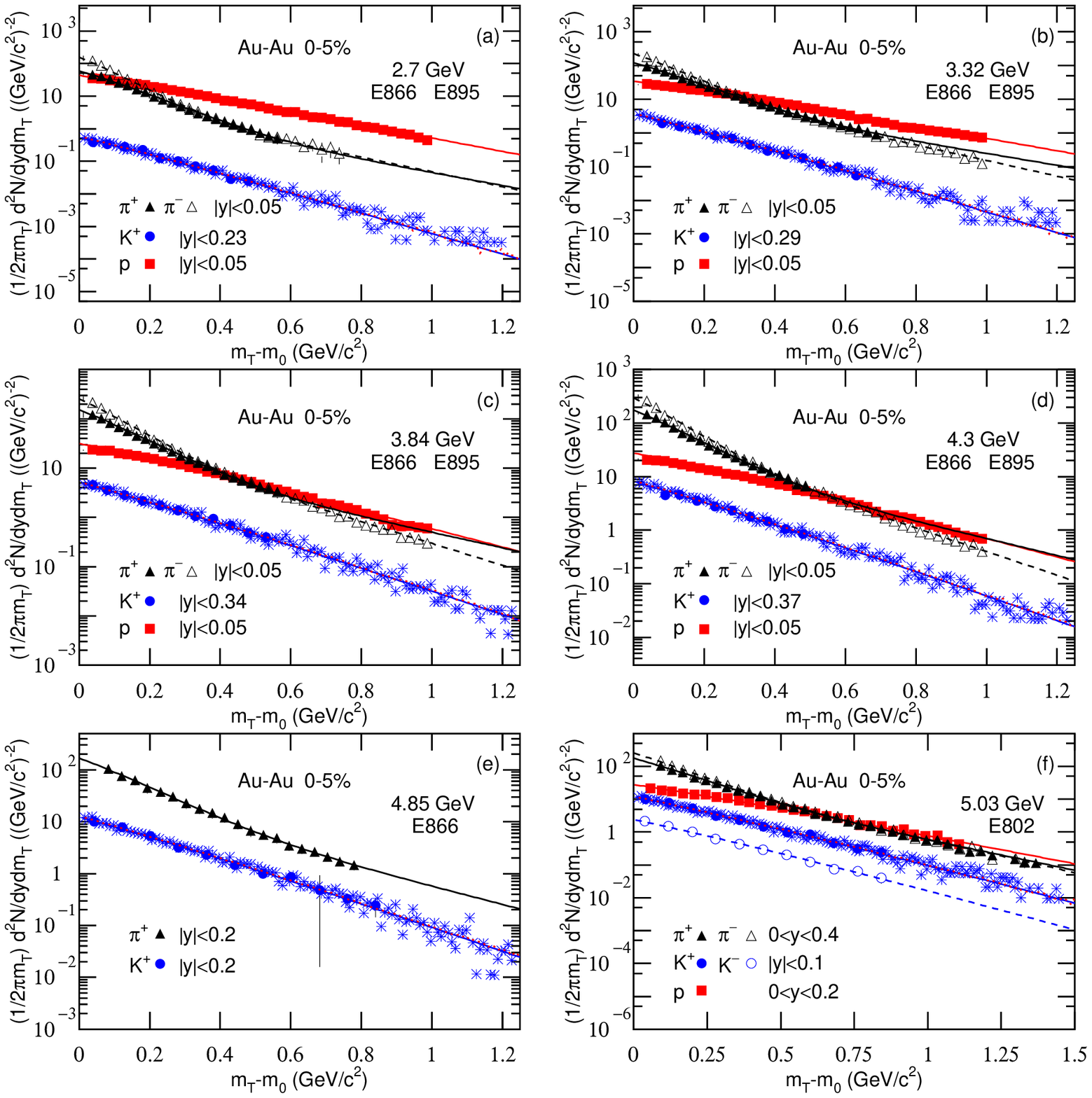}
\end{center}
{\small Fig. 1. The $p_T$ ($m_T-m_0$) spectra, $(1/2\pi p_T)\cdot
d^2N/dydp_T$ [$(1/2\pi m_T)\cdot d^2N/dydm_T$], of $\pi^+$,
$\pi^-$, $K^+$, $K^-$, $p$, and $\bar p$ produced at mid-$y$ or
mid-$\eta$ in central Au-Au collisions at high $\sqrt{s_{NN}}$.
The closed and open symbols represent respectively the
experimental data of positively and negatively charged particles
measured by (a)--(e) the E866~\cite{23} and E895~\cite{24,25}, (f)
the E802~\cite{26,27}, (g)--(o) the STAR~\cite{28,29,30}, and
(p)--(q) the PHENIX~\cite{31,32} Collaborations marked in the
panels which appear mostly [(g)--(q)] in Fig. 1 continued parts,
where in Figs. 1(a)--1(d) the data for $\pi^{\pm}$ and $K^+$ are
taken from the E866 Collaboration~\cite{23} and the data for $p$
are taken from the E895 Collaboration~\cite{24,25}. The solid and
dashed curves are our results fitted by Eq. (6) or (7) for
positively and negatively charged particles respectively. The
dot-dashed curves are our results fitted by using the weighted
average parameter $\langle T\rangle$. The dotted curves and
asterisks in Figs. 1(a)--1(f) represent the MC results for $K^+$
with high ($10^6$ particles) and low ($10^4$ particles) statistics
respectively.}
\end{figure*}

\begin{figure*}[!htb]
\begin{center}
\includegraphics[width=16.0cm]{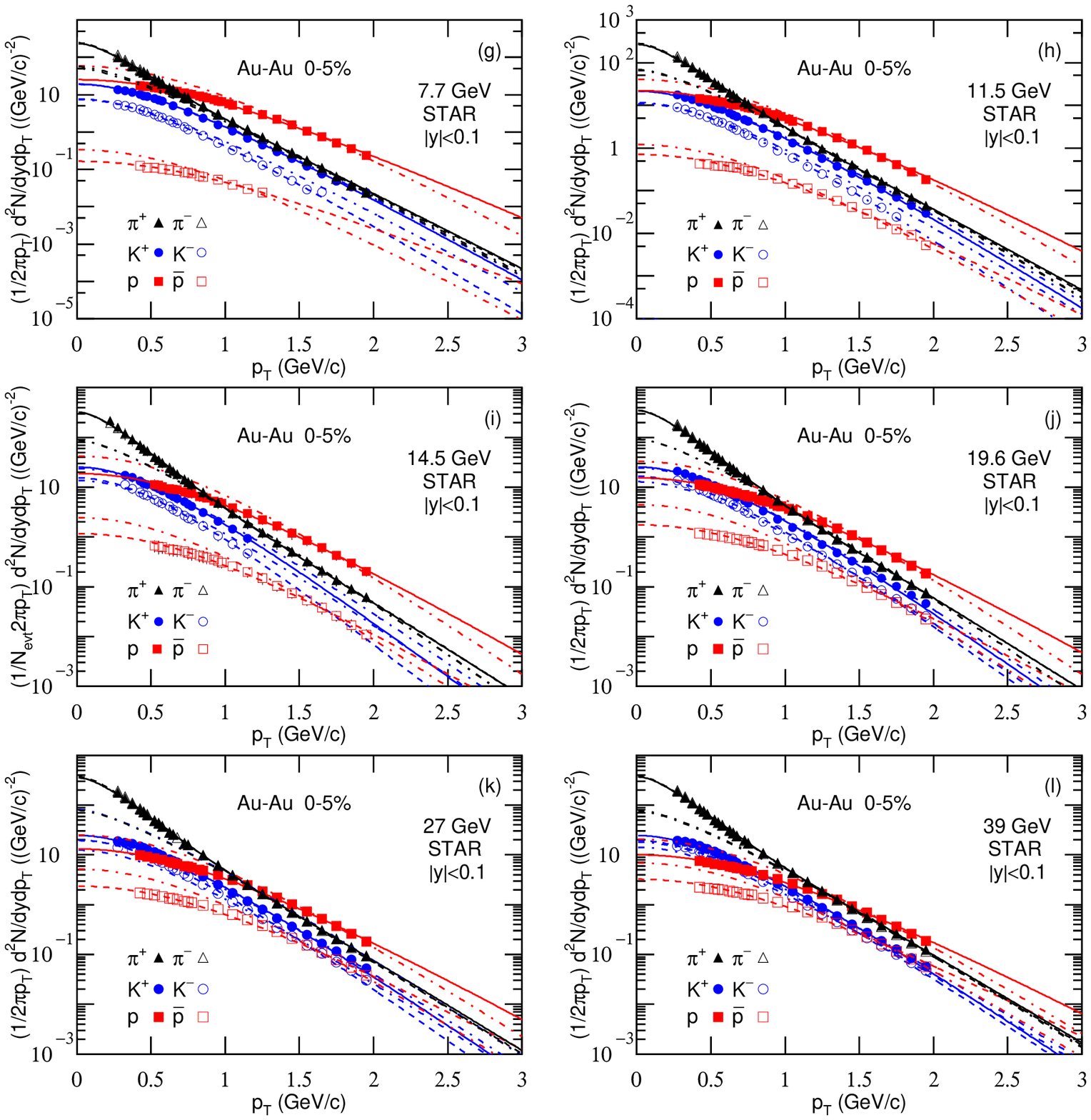}
\end{center}
{\small Fig. 1. Continued.}
\end{figure*}

\begin{figure*}[!htb]
\begin{center}
\includegraphics[width=16.0cm]{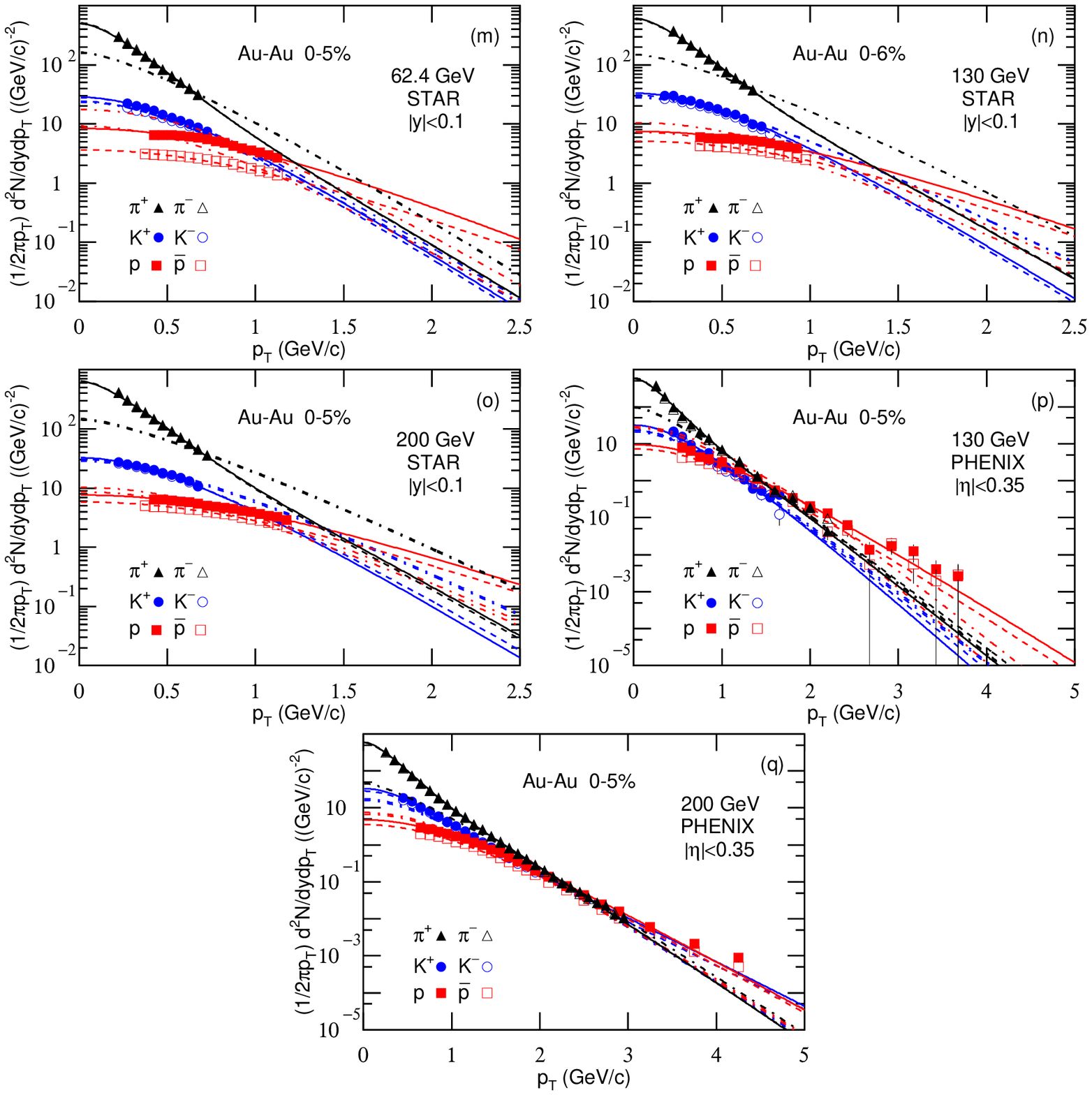}
\end{center}
{\small Fig. 1. Continued.}
\end{figure*}

\begin{figure*}[!htb]
\begin{center}
\includegraphics[width=16.0cm]{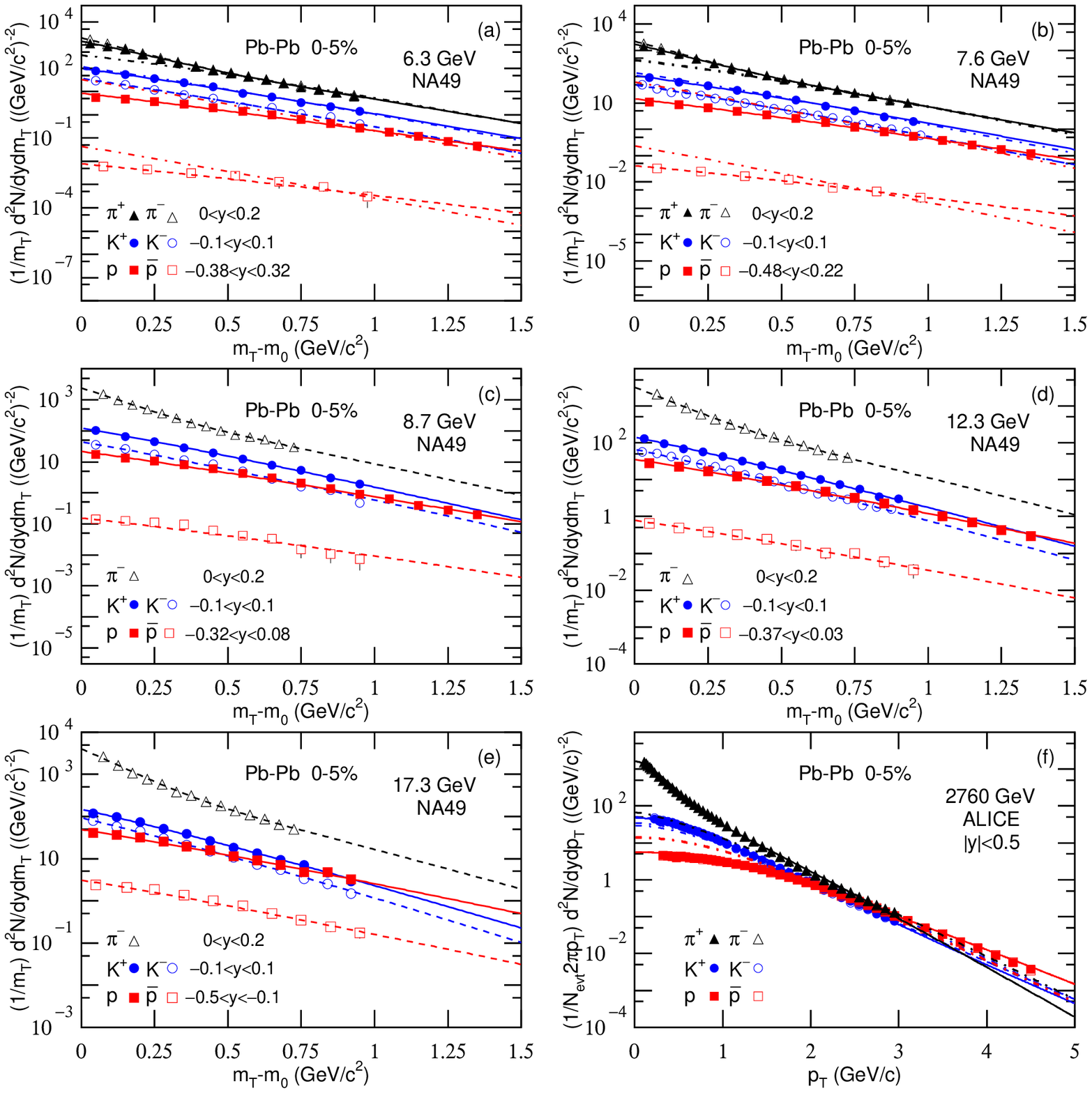}
\end{center}
{\small Fig. 2. Same as Fig. 1, but showing the spectra of various
particles produced at mid-$y$ in central Pb-Pb collisions at high
$\sqrt{s_{NN}}$, where the factor $(1/2\pi)$ on the vertical axis
is removed in some cases and $N_{evt}$ if available denotes the
particle number which can be removed. The symbols represent the
experimental data measured by (a)--(e) the NA49~\cite{33,34,35}
and (f) the ALICE~\cite{36} Collaborations.}
\end{figure*}

{\section{Results and discussion}}

Figures 1(a)--1(q) show the $p_T$ ($m_T-m_0$) spectra, $(1/2\pi
p_T)\cdot d^2N/dydp_T$ [$(1/2\pi m_T)\cdot d^2N/dydm_T$], of
$\pi^+$, $\pi^-$, $K^+$, $K^-$, $p$, and $\bar p$ produced at
mid-$y$ or mid-$\eta$ in central Au-Au collisions at different
$\sqrt{s_{NN}}$, where the particle types, $y$ or $\eta$
intervals, centrality classes, and collision energies are marked
in the panels. The closed and open symbols represent respectively
the experimental data of positively and negatively charged
particles measured by the E866~\cite{23}, E895~\cite{24,25},
E802~\cite{26,27}, STAR~\cite{28,29,30}, and PHENIX~\cite{31,32}
Collaborations marked in the panels, where in Figs. 1(a)--1(d) the
data for $\pi^{\pm}$ and $K^+$ are taken from the E866
Collaboration~\cite{23} and the data for $p$ are taken from the
E895 Collaboration~\cite{24,25}. The solid and dashed curves are
our results fitted by Eq. (6) or (7) for positively and negatively
charged particles respectively. The values of free parameters
($T_1$, $T_2$ if available, $k$), derived parameter ($T$),
normalization constant ($N_0$), $\chi^2$, and degree-of-freedom
(dof) are listed in Table 1. The dot-dashed curves are our results
fitted by using the single component function with the weighted
average parameter $\langle T\rangle$ which will be discussed
later. The dotted curves and asterisks in Figs. 1(a)--1(f)
represent the MC results for $K^+$ with high ($10^6$ particles)
and low ($10^4$ particles) statistics respectively, which will be
discussed at the end of this section. In the fitting process for
the solid and dashed curves, the least squares method is used to
determine the best parameter values. The experimental global
uncertainties used in the calculation of $\chi^2$ are taken to be
the root sum square of statistical uncertainties and
point-by-point systematic uncertainties. The best parameters are
determined due to the limitation of the minimum $\chi^2$. The
global uncertainties of parameters are obtained using the method
of statistical simulation~\cite{59}. We note that $\chi^2$ per dof
($\chi^2$/dof) in a few cases is larger than 10, which renders
that the fit is not too good. One can see that the (two-component)
standard distribution fits approximately the $p_T$ ($m_T-m_0$)
spectra of $\pi^{\pm}$, $K^{\pm}$, $p$, and $\bar p$ measured at
mid-$y$ or mid-$\eta$ in central Au-Au collisions over an energy
range from 2.7 to 200 GeV in most cases.

Figure 2 is similar to Fig. 1, but it showing the spectra of
various particles produced at mid-$y$ in central Pb-Pb collisions
at high $\sqrt{s_{NN}}$, where the factor $(1/2\pi)$ on the
vertical axis is removed in some cases and $N_{evt}$ if available
denotes the particle number which can be removed. The symbols
represent the experimental data measured by the
NA49~\cite{33,34,35} and ALICE~\cite{36} Collaborations. The
values of various parameters, $\chi^2$, and dof for fitting the
solid and dashed curves are listed in Table 2. Figure 3 is similar
to Figs. 1 and 2, but it showing the spectra of various particles
produced at mid-$y$ or mid-$\eta$ in INEL $pp$ collisions at high
center-of-mass energies $\sqrt{s}$, where the factor $(1/2\pi
p_T)$ on the vertical axis is removed and the invariant
cross-section $Ed^3\sigma/dp^3$ is used in some cases. The symbols
represent the experimental data measured by the
NA61/SHINE~\cite{37,38}, PHENIX~\cite{39}, and CMS~\cite{40,41}
Collaborations. The values of various parameters, $\chi^2$, and
dof for fitting the solid and dashed curves are listed in Table 3,
where the cross-section $\sigma_0$ is used as the normalization
constant if the spectrum is $Ed^3\sigma/dp^3$. One can see that
the (two-component) standard distribution fits approximately the
$p_T$ ($m_T-m_0$) spectra of $\pi^{\pm}$, $K^{\pm}$, $p$, and
$\bar p$ measured at mid-$y$ or mid-$\eta$ in central Pb-Pb and
INEL $pp$ collisions over an energy range from 6.3 to 13000 GeV in
most cases.

To study the dependence of main parameter $T$ on collision energy
$\sqrt{s_{NN}}$ or $\sqrt{s}$, the excitation functions of $T$ for
central Au-Au (Pb-Pb) collisions and INEL $pp$ collisions are
shown in Figs. 4(a) and 4(b) respectively. The results obtained
from the spectra of $\pi^+$, $\pi^-$, $K^+$, $K^-$, $p$, and $\bar
p$ are displayed by different symbols marked in the panels. The
asterisks represent $\langle T\rangle$ which is the average $T$ by
weighting different masses and yields of the six particles. In the
case of one of the six particles is absent, $\langle T\rangle$ is
not available. One can see that $T$ and $\langle T\rangle$
increase with the increase of $\ln(\sqrt{s_{NN}})$
[$\ln(\sqrt{s})$]. Meanwhile, $T$ increases with the increase of
particle mass.

\end{multicols}
\renewcommand{\baselinestretch}{1.}
\begin{sidewaystable}
%\vskip-1.cm
%\begin{table*}
{\small Table 1. Values of free parameters ($T_1$, $T_2$ if
available, $k$), derived parameter [$T=kT_1+(1-k)T_2$],
normalization constant ($N_0$), $\chi^2$ and dof corresponding to
the solid and dashed curves in Fig. 1 in which different data in
central Au-Au collisions are measured in different
mid-(pseudo)rapidity intervals at different energies by different
collaborations. The values for positive and negative particles are
listed in terms of value$_1$/value$_2$, if the two values appear
together.
\begin{center}\vskip-4mm
\scriptsize
\begin{tabular} {cccccccccc}\\ \hline\hline Collab. & $\sqrt{s_{NN}}$ (GeV) &
Particle & $T_1$ (MeV) & $T_2$ (MeV) & $k$  & $T$ (MeV) & $N_0$ & $\chi^2$ & dof \\
\hline
E866/E895 & 2.7  & $\pi^\pm$ & $80\pm4/62\pm4$   & $190\pm5/165\pm5$  & $0.88\pm0.02/0.88\pm0.02$ & $93\pm4/74\pm4$   & $1.19\pm0.03/2.08\pm0.04$   & 15/22 & 19/26\\
          &~     & $K^+$     & $128\pm4$         & $-$                & $1$                       & $128\pm4$         & $0.16\pm0.01$               & 4     & 8\\
          &~     & $p$       & $192\pm6$         & $-$                & $1$                       & $192\pm3$         & $1.22\pm0.02$               & 278   & 38\\
          & 3.32 & $\pi^\pm$ & $85\pm4/71\pm4$   & $209\pm12/163\pm8$ & $0.78\pm0.03/0.79\pm0.02$ & $112\pm4/90\pm4$  & $2.84\pm0.02/3.96\pm0.04$   & 8/32  & 24/36\\
          &~     & $K^+$     & $131\pm5$         & $-$                & $1$                       & $131\pm5$         & $1.35\pm0.02$               & 2     & 10\\
          &~     & $p$       & $213\pm4$         & $-$                & $1$                       & $213\pm4$         & $6.99\pm0.05$               & 401   & 38\\
          & 3.84 & $\pi^\pm$ & $88\pm3/75\pm3$   & $230\pm5/170\pm5$  & $0.74\pm0.03/0.73\pm0.02$ & $125\pm3/101\pm3$ & $4.00\pm0.03/5.16\pm0.04$   & 6/54  & 19/36\\
          &~     & $K^+$     & $167\pm5$         & $-$                & $1$                       & $167\pm5$         & $3.08\pm0.01$               & 1     & 9\\
          &~     & $p$       & $210\pm4$         & $-$                & $1$                       & $210\pm4$         & $6.32\pm0.05$               & 592   & 38\\
          & 4.3  & $\pi^\pm$ & $90\pm4/77\pm4$   & $229\pm4/171\pm3$  & $0.70\pm0.02/0.71\pm0.01$ & $132\pm3/104\pm3$ & $4.88\pm0.03/6.30\pm0.04$   & 10/50 & 16/36\\
          &~     & $K^+$     & $172\pm5$         & $-$                & $1$                       & $172\pm5$         & $5.59\pm0.06$               & 1     & 7\\
          &~     & $p$       & $224\pm5$         & $-$                & $1$                       & $224\pm5$         & $6.09\pm0.07$               & 204   & 38\\
          & 4.85 & $\pi^+$   & $96\pm3$          & $214\pm5$          & $0.73\pm0.03$             & $128\pm3$         & $319.96\pm1.12$             & 1     & 16\\
          &~     & $K^+$     & $170\pm5$         & $-$                & $1$                       & $170\pm5$         & $4.66\pm0.10$               & 2     & 9\\
\hline
E802      & 5.03 & $\pi^\pm$ & $103\pm3/93\pm4$  & $210\pm4/194\pm5$  & $0.73\pm0.03/0.72\pm0.02$ & $132\pm2/121\pm3$ & $22.00\pm3.00/27.12\pm3.00$ & 17/45 & 30/29\\
          &~     & $K^\pm$   & $172\pm5/166\pm5$ & $-/-$              & $1/1$                     & $172\pm5/166\pm5$ & $2.28\pm0.05/0.45\pm0.01$   & 3/4   & 9/9\\
          &~     & $p$       & $230\pm4$         & $-$                & $1$                       & $230\pm4$         & $12.59\pm1.03$              & 136   & 27\\
\hline
STAR   & 7.7  & $\pi^\pm$   & $115\pm10/115\pm10$ & $199\pm3/198\pm3$   & $0.68\pm0.02/0.70\pm0.02$ & $142\pm10/140\pm9$  & $18.21\pm3.00/18.83\pm4.00$     & 14/16  & 22/22\\
       &~     & $K^\pm$     & $185\pm5/170\pm5$   & $-/-$               & $1/1$                     & $185\pm5/170\pm5$   & $4.19\pm0.50/1.49\pm0.30$       & 11/12  & 21/21\\
       &~     & $p/\bar{p}$ & $226\pm7/251\pm6$   & $-/-$               & $1/1$                     & $226\pm7/251\pm6$   & $11.01\pm0.70/0.084\pm0.002$    & 11/3   & 27/13\\
       & 11.5 & $\pi^\pm$   & $124\pm10/123\pm10$ & $211\pm3/209\pm3$   & $0.72\pm0.02/0.73\pm0.02$ & $148\pm9/146\pm9$   & $22.82\pm0.50/23.58\pm0.70$     & 8/10   & 22/22\\
       &~     & $K^\pm$     & $190\pm5/178\pm5$   & $-/-$               & $1/1$                     & $190\pm5/178\pm5$   & $4.93\pm0.20/2.43\pm0.10$       & 7/4    & 23/21\\
       &~     & $p/\bar{p}$ & $223\pm7/223\pm7$   & $-/-$               & $1/1$                     & $223\pm7/223\pm7$   & $9.41\pm0.60/0.30\pm0.01$       & 13/32  & 26/21\\
       & 14.5 & $\pi^\pm$   & $124\pm10/123\pm10$ & $213\pm3/212\pm3$   & $0.72\pm0.02/0.70\pm0.02$ & $149\pm9/150\pm9$   & $27.58\pm1.30/27.38\pm1.20$     & 1/1    & 24/24\\
       &~     & $K^\pm$     & $183\pm9/173\pm8$   & $-/-$               & $1/1$                     & $183\pm9/173\pm8$   & $5.49\pm0.03/3.07\pm0.02$       & 1/1    & 16/16\\
       &~     & $p/\bar{p}$ & $230\pm9/229\pm7$   & $-/-$               & $1/1$                     & $230\pm9/229\pm7$   & $8.39\pm0.06/0.52\pm0.01$       & 1/1    & 23/23\\
       & 19.6 & $\pi^\pm$   & $124\pm10/123\pm10$ & $216\pm3/216\pm3$   & $0.70\pm0.03/0.70\pm0.02$ & $152\pm9/151\pm9$   & $29.98\pm5.10/30.00\pm5.40$     & 6/8    & 22/21\\
       &~     & $K^\pm$     & $193\pm6/189\pm5$   & $-/-$               & $1/1$                     & $193\pm6/189\pm5$   & $5.99\pm0.05/3.77\pm0.02$       & 35/36  & 24/24\\
       &~     & $p/\bar{p}$ & $237\pm8/243\pm9$   & $-/-$               & $1/1$                     & $237\pm8/243\pm9$   & $7.17\pm0.07/0.85\pm0.01$       & 6/15   & 27/20\\
       & 27   & $\pi^\pm$   & $125\pm10/124\pm10$ & $220\pm3/217\pm3$   & $0.68\pm0.02/0.69\pm0.02$ & $155\pm10/153\pm9$  & $31.86\pm4.10/32.86\pm4.50$     & 5/10   & 21/22\\
       &~     & $K^\pm$     & $199\pm6/190\pm6$   & $-/-$               & $1/1$                     & $199\pm6/190\pm6$   & $6.07\pm0.05/4.57\pm0.03$       & 19/32  & 24/22\\
       &~     & $p/\bar{p}$ & $242\pm10/251\pm12$ & $-/-$               & $1/1$                     & $242\pm10/251\pm12$ & $6.31\pm0.07/1.20\pm0.01$       & 6/15   & 21/20\\
       & 39 & $\pi^\pm$     & $125\pm10/125\pm10$ & $225\pm3/224\pm3$   & $0.66\pm0.02/0.68\pm0.01$ & $159\pm9/157\pm9$   & $33.80\pm5.10/34.12\pm5.30$     & 5/6    & 22/22\\
       &~     & $K^\pm$     & $203\pm7/203\pm6$   & $-/-$               & $1/1$                     & $203\pm7/203\pm6$   & $6.23\pm0.07/4.91\pm0.03$       & 16/28  & 24/24\\
       &~     & $p/\bar{p}$ & $257\pm11/259\pm10$ & $-/-$               & $1/1$                     & $257\pm11/259\pm10$ & $4.91\pm0.05/1.72\pm0.01$       & 5/19   & 20/21\\
       & 62.4 & $\pi^\pm$   & $123\pm10/123\pm10$ & $220\pm3/220\pm3$   & $0.71\pm0.01/0.72\pm0.01$ & $151\pm9/150\pm9$   & $42.30\pm5.12/43.12\pm5.20$     & 22/25  & 6/6\\
       &~     & $K^\pm$     & $210\pm8/210\pm8$   & $-/-$               & $1/1$                     & $210\pm8/210\pm8$   & $7.61\pm0.30/6.50\pm0.27$       & 1/10   & 8/8\\
       &~     & $p/\bar{p}$ & $320\pm11/346\pm11$ & $-/-$               & $1/1$                     & $320\pm11/346\pm11$ & $6.25\pm0.20/3.10\pm0.05$       & 170/102& 13/14\\
       & 130  & $\pi^\pm$   & $120\pm6/123\pm5$   & $232\pm9/234\pm10$  & $0.68\pm0.01/0.69\pm0.01$ & $156\pm6/157\pm5$   & $51.86\pm5.22/51.92\pm5.66$     & 39/74  & 6/6\\
       &~     & $K^\pm$     & $216\pm8/214\pm9$   & $-/-$               & $1/1$                     & $216\pm8/214\pm9$   & $9.32\pm0.28/8.48\pm0.15$       & 2/3    & 10/11\\
       &~     & $p/\bar{p}$ & $353\pm9/359\pm8$   & $-/-$               & $1/1$                     & $353\pm9/359\pm8$   & $6.64\pm0.15/4.68\pm0.11$       & 57/134 & 10/11\\
       & 200  & $\pi^\pm$   & $126\pm9/127\pm10$  & $236\pm12/235\pm12$ & $0.67\pm0.01/0.70\pm0.01$ & $162\pm9/159\pm10$  & $59.00\pm3.33/59.12\pm3.70$     & 18/23  & 7/7\\
       &~     & $K^\pm$     & $221\pm13/230\pm13$ & $-/-$               & $1/1$                     & $221\pm13/230\pm13$ & $9.48\pm0.23/9.50\pm0.25$       & 3/2    & 8/8\\
       &~     & $p/\bar{p}$ & $376\pm16/372\pm11$ & $-/-$               & $1/1$                     & $376\pm16/372\pm11$ & $7.56\pm0.18/5.68\pm0.14$       & 4/31   & 14/15\\
\hline
PHENIX & 130  & $\pi^\pm$   & $123\pm10/121\pm10$ & $220\pm3/229\pm3$   & $0.70\pm0.02/0.73\pm0.01$ & $152\pm9/150\pm10$  & $167.72\pm10.50/167.72\pm11.00$ & 146/182& 10/10\\
       &~     & $K^\pm$     & $201\pm10/213\pm13$ & $-/-$               & $1/1$                     & $210\pm10/213\pm13$ & $27.31\pm2.10/21.36\pm1.53$     & 17/15  & 11/11\\
       &~     & $p/\bar{p}$ & $276\pm10/267\pm11$ & $-/-$               & $1/1$                     & $276\pm10/267\pm11$ & $18.77\pm1.25/13.80\pm1.16$     & 15/13  & 15/15\\
       & 200  & $\pi^\pm$   & $136\pm4/136\pm4$   & $259\pm3/259\pm3$   & $0.74\pm0.02/0.73\pm0.02$ & $168\pm4/169\pm4$   & $199.02\pm16.01/196.43\pm15.10$ & 286/205& 24/24\\
       &~     & $K^\pm$     & $197\pm12/215\pm13$ & $339\pm11/346\pm12$ & $0.73\pm0.01/0.81\pm0.01$ & $235\pm12/240\pm13$ & $32.56\pm2.10/30.04\pm2.00$     & 28/22  & 12/12\\
       &~     & $p/\bar{p}$ & $311\pm7/314\pm8$   & $-/-$               & $1/1$                     & $311\pm7/314\pm8$   & $11.49\pm0.72/8.62\pm0.06$      & 69/90  & 20/20\\
\hline
\end{tabular}%
\end{center}}
%\end{table*}
\end{sidewaystable}

\begin{sidewaystable}
%\vskip-1.cm
%\begin{table*}
{\small Table 2. The same as Table 1, but showing the results for
central Pb-Pb collisions which are shown in Fig. 2.
\begin{center}\vskip-4mm
\scriptsize
\begin{tabular} {cccccccccc}\\ \hline\hline Collab. & $\sqrt{s_{NN}}$ (GeV) &
Particle & $T_1$ (MeV) & $T_2$ (MeV) & $k$  & $T$ (MeV) & $N_0$ & $\chi^2$ & dof \\
\hline
NA49  & 6.3  & $\pi^\pm$   & $97\pm5/86\pm4$    & $185\pm5/181\pm4$ & $0.64\pm0.02/0.65\pm0.01$& $129\pm5/119\pm4$  & $14.50\pm1.10/16.98\pm1.23$    & 69/43 & 12/12\\
      &~     & $K^\pm$     & $182\pm5/172\pm8$  & $-/-$             & $1/1$                    & $182\pm5/172\pm8$  & $3.33\pm0.03/1.14\pm0.01$      & 22/25 & 8/8\\
      &~     & $p/\bar{p}$ & $227\pm10/259\pm12$& $-/-$             & $1/1$                    & $227\pm10/259\pm12$& $2.05\pm0.01/0.0023\pm0.0001$  & 14/2  & 12/5\\
      & 7.6  & $\pi^\pm$   & $97\pm6/86\pm4$    & $195\pm4/187\pm3$ & $0.61\pm0.01/0.58\pm0.01$& $135\pm6/128\pm4$  & $17.44\pm1.15/19.50\pm2.20$    & 24/26 & 12/12\\
      &~     & $K^\pm$     & $193\pm5/181\pm6$  & $-/-$             & $1/1$                    & $193\pm5/181\pm6$  & $4.11\pm0.33/1.63\pm0.05$      & 30/60 & 8/18\\
      &~     & $p/\bar{p}$ & $242\pm6/284\pm11$ & $-/-$             & $1/1$                    & $242\pm6/284\pm11$ & $3.74\pm0.08/0.014\pm0.001$    & 10/1  & 12/5\\
      & 8.7  & $\pi^-$     & $84\pm5$           & $188\pm4$         & $0.56\pm0.02$            & $130\pm5$          & $21.40\pm0.71$                 & 44    & 10\\
      &~     & $K^\pm$     & $185\pm6/187\pm5$  & $-/-$             & $1/1$                    & $185\pm6/187\pm5$  & $4.30\pm0.05/1.58\pm0.01$      & 37/33 & 8/8\\
      &~     & $p/\bar{p}$ & $244\pm7/280\pm10$ & $-/-$             & $1/1$                    & $244\pm7/280\pm10$ & $3.56\pm0.03/0.030\pm0.001$    & 14/10 & 12/8\\
      & 12.3 & $\pi^-$     & $84\pm6$           & $189\pm5$         & $0.56\pm0.01$            & $130\pm6$          & $27.80\pm2.11$                 & 60    & 10\\
      &~     & $K^\pm$     & $185\pm7/183\pm5$  & $-/-$             & $1/1$                    & $185\pm7/183\pm5$  & $4.92\pm0.08/2.24\pm0.02$      & 10/54 & 12/16\\
      &~     & $p/\bar{p}$ & $244\pm8/260\pm14$ & $-/-$             & $1/1$                    & $244\pm8/260\pm14$ & $5.48\pm0.09/0.13\pm0.01$      & 5/1   & 12/8\\
      & 17.3 & $\pi^-$     & $84\pm6$           & $200\pm5$         & $0.57\pm0.01$            & $134\pm6$          & $35.80\pm1.80$                 & 47    & 10\\
      &~     & $K^\pm$     & $193\pm7/185\pm6$  & $-/-$             & $1/1$                    & $193\pm7/185\pm6$  & $5.52\pm0.30/3.26\pm0.09$      & 27/17 & 10/10\\
      &~     & $p/\bar{p}$ & $280\pm13/280\pm14$& $-/-$             & $1/1$                    & $280\pm13/280\pm14$& $8.86\pm0.23/0.56\pm0.03$      & 9/3   & 10/8\\
\hline
ALICE & 2760 & $\pi^\pm$   & $127\pm4/127\pm4$  & $309\pm3/309\pm3$ & $0.65\pm0.01/0.65\pm0.01$& $191\pm4/191\pm4$  & $762.61\pm33.11/762.61\pm33.11$& 74/73 & 37/37\\
      &~     & $K^\pm$     & $290\pm6/275\pm4$  & $409\pm10/409\pm3$& $0.82\pm0.02/0.75\pm0.02$& $311\pm6/309\pm4$  & $108.82\pm5.3/108.49\pm5.1$    & 49/4  & 32/32\\
      &~     & $p/\bar{p}$ & $428\pm9/428\pm6$  & $-/-$             & $1/1$                    & $428\pm9/428\pm6$  & $32.81\pm2.10/32.91\pm1.91$    & 63/49 & 40/40\\
 \hline
\end{tabular}%
\end{center}}
%\end{table*}
%\end{sidewaystable}

%\begin{sidewaystable}
%\vskip-1.cm
%\begin{table*}
{\small Table 3. The same as Table 1, but showing the results for
INEL $pp$ collisions which are shown in Fig. 3, where the
normalization constant is $\sigma_0$ for the cases of 62.4 and 200
GeV presented in Figs. 3(f) and 3(g) respectively.
\begin{center}\vskip-4mm
\scriptsize
\begin{tabular} {cccccccccc}\\ \hline\hline Collab. & $\sqrt{s}$ (GeV) &
Particle & $T_1$ (MeV) & $T_2$ (MeV) & $k$  & $T$ (MeV) & $N_0$ ($\sigma_0$) & $\chi^2$ & dof \\
\hline
NA61/SHINE & 6.3  & $\pi^-$     & $110\pm7$           & $169\pm9$          & $0.82\pm0.01$             & $121\pm7$           & $0.086\pm0.006$                  & 12   & 14\\
           &~     & $K^\pm$     & $113\pm6/123\pm6$   & $-/-$              & $1/1$                     & $113\pm6/123\pm6$   & $0.0096\pm0.0003/0.0030\pm0.0001$& 2/1  & 4/8\\
           &~     & $p$         & $125\pm5$           & $-$                & $1$                       & $125\pm5$           & $0.027\pm0.001$                  & 69   & 5\\
           & 7.7  & $\pi^\pm$   & $103\pm10/113\pm10$ & $177\pm9/178\pm9$  & $0.78\pm0.01/0.81\pm0.01$ & $119\pm10/125\pm10$ & $0.16\pm0.01/0.098\pm0.003$      & 1/16 & 1/16\\
           &~     & $K^\pm$     & $131\pm6/127\pm6$   & $-/-$              & $1/1$                     & $131\pm6/127\pm6$   & $0.0013\pm0.0001/0.0050\pm0.0002$& 4/1  & 7/5\\
           &~     & $p/\bar{p}$ & $136\pm10/104\pm12$ & $-/-$              & $1/1$                     & $136\pm10/104\pm12$ & $0.027\pm0.002/0.00066\pm0.00001$& 4/5  & 9/7\\
           & 8.8  & $\pi^\pm$   & $105\pm10/113\pm10$ & $182\pm9/178\pm9$  & $0.78\pm0.02/0.81\pm0.02$ & $122\pm10/125\pm10$ & $0.16\pm0.01/0.11\pm0.01$        & 1/33 & 2/18\\
           &~     & $K^\pm$     & $126\pm6/123\pm6$   & $-/-$              & $1/1$                     & $126\pm6/123\pm6$   & $0.013\pm0.001/0.0056\pm0.0002$  & 24/6 & 9/7\\
           &~     & $p/\bar{p}$ & $136\pm9/102\pm12$  & $-/-$              & $1/1$                     & $136\pm9/102\pm12$  & $0.027\pm0.001/0.00078\pm0.00004$& 1/3  & 9/3\\
           & 12.3 & $\pi^\pm$   & $108\pm10/113\pm10$ & $190\pm9/184\pm9$  & $0.77\pm0.02/0.79\pm0.01$ & $127\pm10/128\pm10$ & $0.16\pm0.01/0.12\pm0.01$        & 1/23 & 2/16\\
           &~     & $K^\pm$     & $135\pm6/127\pm6$   & $-/-$              & $1/1$                     & $135\pm6/127\pm6$   & $0.015\pm0.001/0.0074\pm0.0003$  & 10/9 & 8/7\\
           &~     & $p/\bar{p}$ & $150\pm10/106\pm12$ & $-/-$              & $1/1$                     & $150\pm10/106\pm12$ & $0.017\pm0.001/0.0018\pm0.0001$  & 6/9  & 11/9\\
           & 17.3 & $\pi^\pm$   & $111\pm8/113\pm10$  & $180\pm9/190\pm8$  & $0.76\pm0.01/0.79\pm0.01$ & $128\pm8/129\pm10$  & $0.16\pm0.01/0.13\pm0.01$        & 1/8  & 1/19\\
           &~     & $K^\pm$     & $130\pm6/138\pm6$   & $-/-$              & $1/1$                     & $130\pm6/138\pm6$   & $0.015\pm0.001/0.010\pm0.001$    & 2/2  & 8/8\\
           &~     & $p/\bar{p}$ & $150\pm10/128\pm12$ & $-/-$              & $1/1$                     & $50\pm10/128\pm12$  & $0.015\pm0.001/0.0034\pm0.0001$  & 2/4  & 8/8\\
\hline
PHENIX     & 62.4 & $\pi^\pm$   & $118\pm5/113\pm4$   & $240\pm4/229\pm4$  & $0.82\pm0.01/0.81\pm0.01$ & $140\pm5/135\pm4$   & $17.21\pm3.00/18.40\pm5.00$& 14/25 & 22/22\\
           &~     & $K^\pm$     & $147\pm6/149\pm5$   & $284\pm6/278\pm5$  & $0.76\pm0.01/0.78\pm0.01$ & $180\pm6/177\pm5$   & $1.84\pm0.50/1.58\pm0.40$  & 5/2   & 12/12\\
           &~     & $p/\bar{p}$ & $161\pm10/163\pm12$ & $283\pm4/278\pm5$  & $0.80\pm0.00/0.81\pm0.01$ & $182\pm10/185\pm11$ & $1.05\pm0.40/0.57\pm0.03$  & 12/14 & 23/23\\
           & 200  & $\pi^\pm$   & $117\pm5/115\pm5$   & $278\pm5/273\pm5$  & $0.85\pm0.00/0.82\pm0.01$ & $141\pm5/145\pm5$   & $25.26\pm5.00/22.61\pm3.00$& 101/83& 23/23\\
           &~     & $K^\pm$     & $151\pm4/149\pm5$   & $303\pm8/301\pm5$  & $0.72\pm0.01/0.72\pm0.01$ & $193\pm4/192\pm5$   & $2.78\pm0.40/2.76\pm0.70$  & 10/7  & 12/12\\
           &~     & $p/\bar{p}$ & $162\pm5/160\pm6$   & $316\pm7/312\pm5$  & $0.73\pm0.00/0.76\pm0.00$ & $204\pm5/197\pm6$   & $1.23\pm0.20/1.09\pm0.10$  & 108/91& 30/30\\
\hline
CMS        & 900  & $\pi^\pm$   & $115\pm4/117\pm5$   & $264\pm5/270\pm6$  & $0.68\pm0.01/0.68\pm0.01$ & $163\pm4/165\pm5$   & $3.76\pm0.50/3.70\pm0.50$  & 13/12 & 18/18\\
           &~     & $K^\pm$     & $156\pm6/154\pm6$   & $343\pm10/340\pm10$& $0.59\pm0.01/0.58\pm0.01$ & $233\pm6/232\pm6$   & $0.47\pm0.03/0.46\pm0.03$  & 4/6   & 13/13\\
           &~     & $p/\bar{p}$ & $177\pm6/177\pm5$   & $331\pm9/328\pm8$  & $0.55\pm0.01/0.56\pm0.01$ & $246\pm6/243\pm5$   & $0.21\pm0.02/0.20\pm0.01$  & 19/17 & 23/23\\
           & 2760 & $\pi^\pm$   & $116\pm3/116\pm4$   & $275\pm5/281\pm5$  & $0.64\pm0.01/0.65\pm0.01$ & $173\pm3/174\pm4$   & $4.68\pm0.90/4.60\pm0.80$  & 19/14 & 18/18\\
           &~     & $K^\pm$     & $165\pm7/164\pm8$   & $351\pm9/350\pm9$  & $0.56\pm0.01/0.56\pm0.01$ & $247\pm7/246\pm8$   & $0.60\pm0.03/0.59\pm0.03$  & 5/10  & 13/13\\
           &~     & $p/\bar{p}$ & $182\pm6/183\pm7$   & $406\pm8/392\pm7$  & $0.54\pm0.01/0.55\pm0.01$ & $285\pm6/277\pm7$   & $0.27\pm0.01/0.26\pm0.01$  & 28/32 & 23/23\\
           & 7000 & $\pi^\pm$   & $116\pm4/115\pm4$   & $292\pm5/288\pm4$  & $0.63\pm0.02/0.62\pm0.01$ & $181\pm4/181\pm4$   & $5.82\pm0.80/5.76\pm0.80$  & 34/22 & 18/18\\
           &~     & $K^\pm$     & $178\pm6/175\pm6$   & $420\pm10/393\pm10$& $0.59\pm0.02/0.56\pm0.02$ & $277\pm6/271\pm6$   & $0.76\pm0.03/0.75\pm0.03$  & 3/8   & 13/13\\
           &~     & $p/\bar{p}$ & $217\pm7/202\pm7$   & $429\pm9/430\pm9$  & $0.57\pm0.02/0.53\pm0.01$ & $308\pm7/309\pm7$   & $0.34\pm0.02/0.33\pm0.02$  & 28/22 & 23/23\\
           & 13000& $\pi^\pm$   & $114\pm4/114\pm6$   & $300\pm5/291\pm6$  & $0.61\pm0.01/0.60\pm0.01$ & $187\pm4/185\pm6$   & $5.46\pm0.70/5.38\pm0.60$  & 9/10  & 18/18\\
           &~     & $K^\pm$     & $180\pm10/180\pm10$ & $360\pm12/383\pm12$& $0.59\pm0.01/0.60\pm0.01$ & $254\pm10/261\pm10$ & $0.65\pm0.03/0.65\pm0.03$  & 3/2   & 13/13\\
           &~     & $p/\bar{p}$ & $229\pm10/231\pm11$ & $425\pm3/425\pm9$  & $0.63\pm0.01/0.62\pm0.01$ & $302\pm10/305\pm10$ & $0.30\pm0.02/0.29\pm0.02$  & 25/19 & 22/22\\
\hline
\end{tabular}%
\end{center}}
%\end{table*}
\end{sidewaystable}
\renewcommand{\baselinestretch}{1.15}
\begin{multicols}{2}

\begin{figure*}[!htb]
\begin{center}
\includegraphics[width=16.0cm]{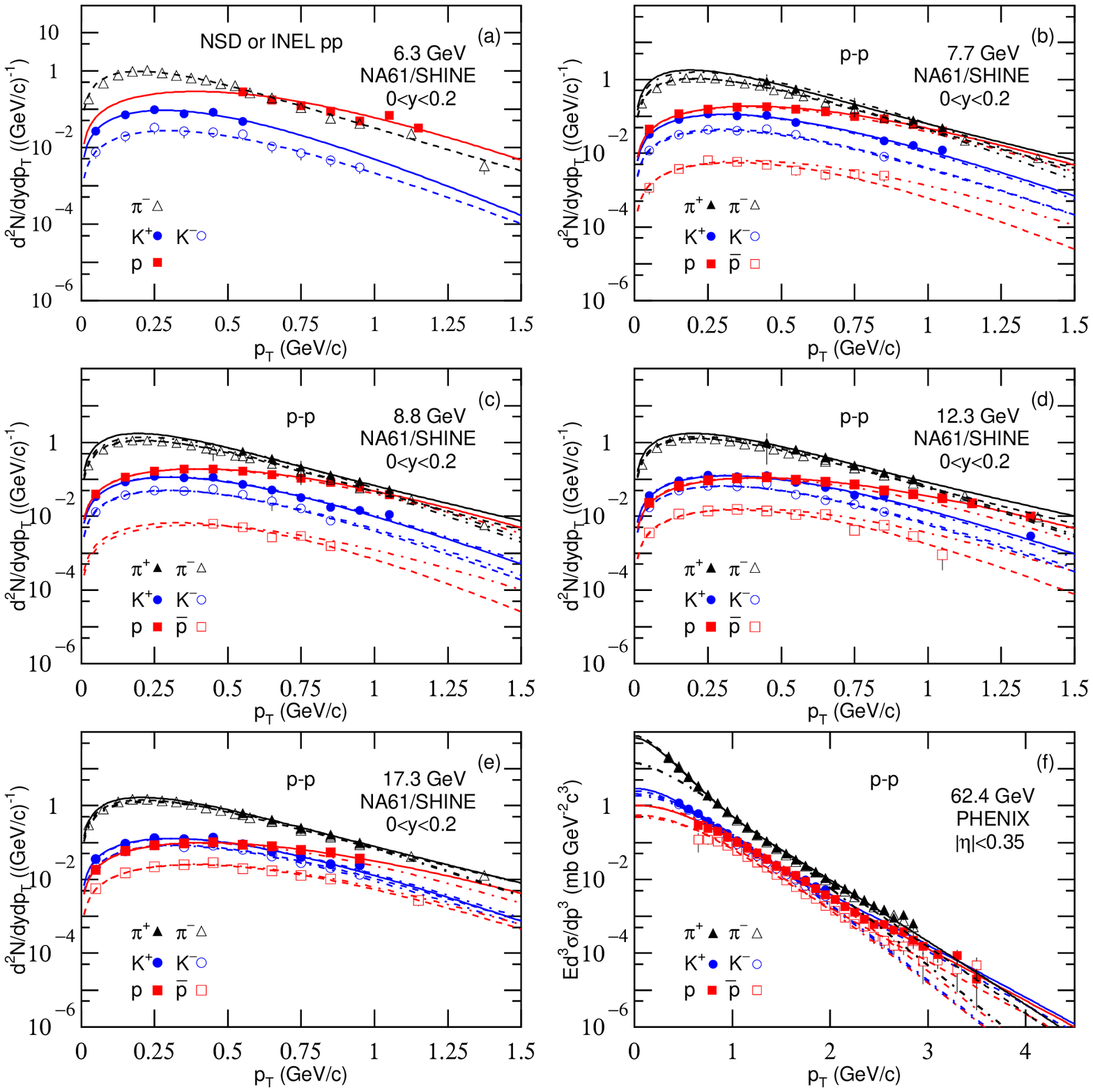}
\end{center}
{\small Fig. 3. Same as Fig. 1, but showing the spectra of various
particles produced at mid-$y$ or mid-$\eta$ in INEL $pp$
collisions at high center-of-mass energies $\sqrt{s}$, where the
factor $(1/2\pi p_T)$ on the vertical axis is removed and the
invariant cross-section $Ed^3\sigma/dp^3$ is used in some cases.
The symbols represent the experimental data measured by (a)--(e)
the NA61/SHINE~\cite{37,38}, (f)--(g) the PHENIX~\cite{39}, and
(h)--(k) the CMS~\cite{40,41} Collaborations, where panels
(g)--(k) appear in Fig. 3 continued part.}
\end{figure*}

\begin{figure*}[!htb]
\begin{center}
\includegraphics[width=16.0cm]{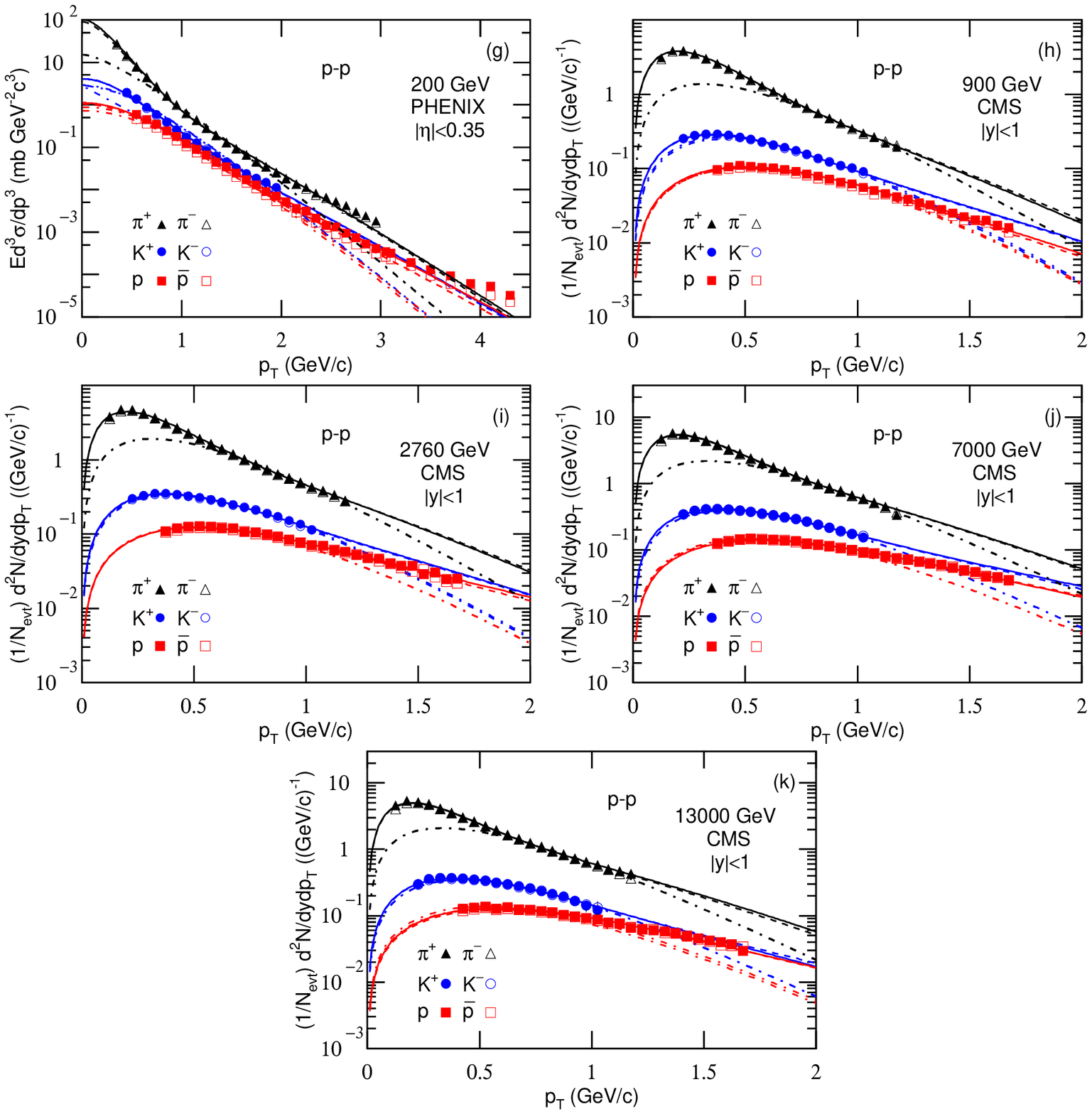}
\end{center}
{\small Fig. 3. Continued.}
\end{figure*}

\begin{figure*}[!htb]
\begin{center}
\includegraphics[width=16.0cm]{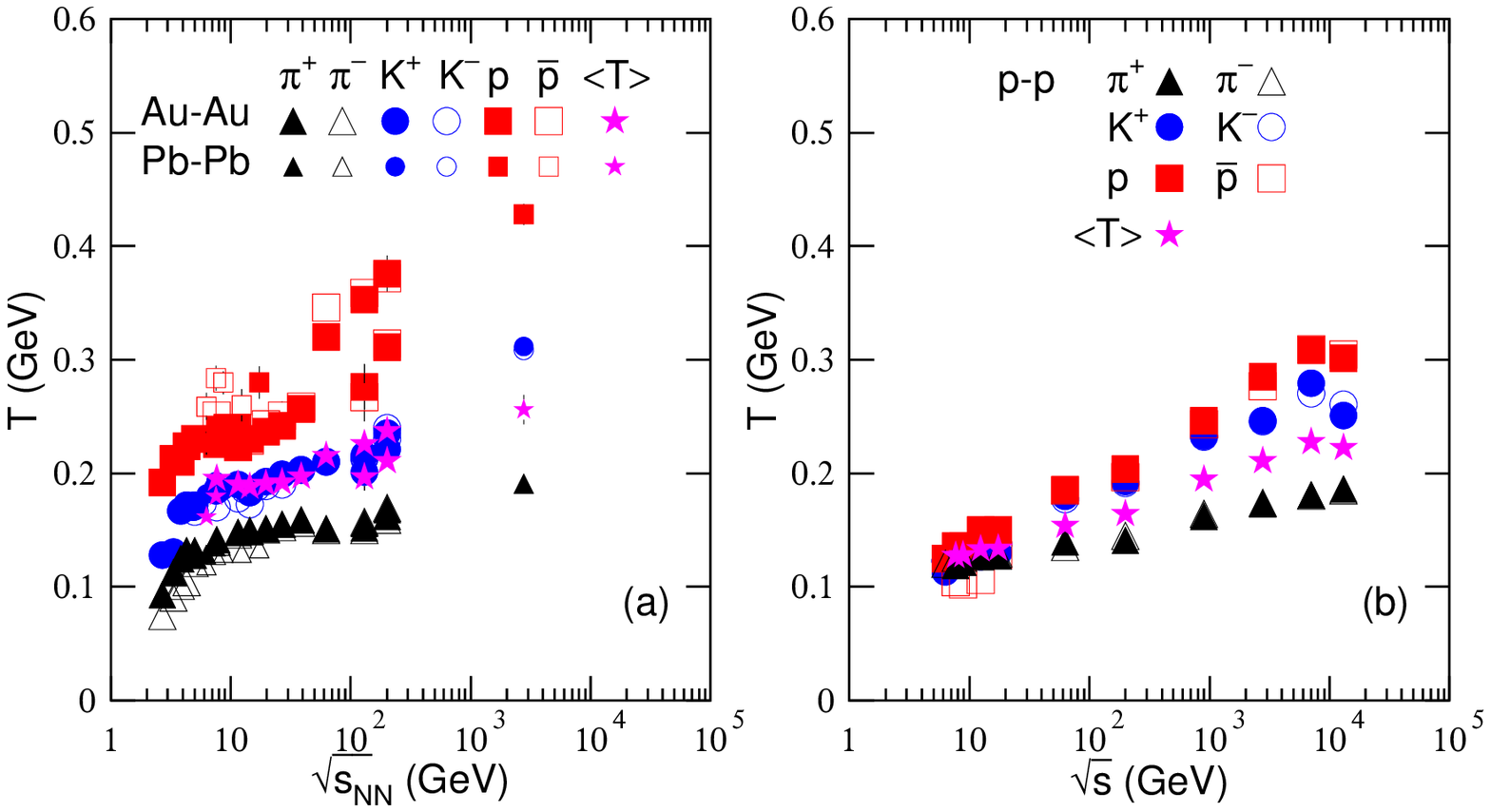}
\end{center}
{\small Fig. 4. Excitation functions of $T$ and $\langle T\rangle$
in (a) central Au-Au (Pb-Pb) collisions and (b) INEL $pp$
collisions. The symbols corresponding to identified particles are
extracted from experimental spectra.}
\end{figure*}

\begin{figure*}[!htb]
\begin{center}
\includegraphics[width=16.0cm]{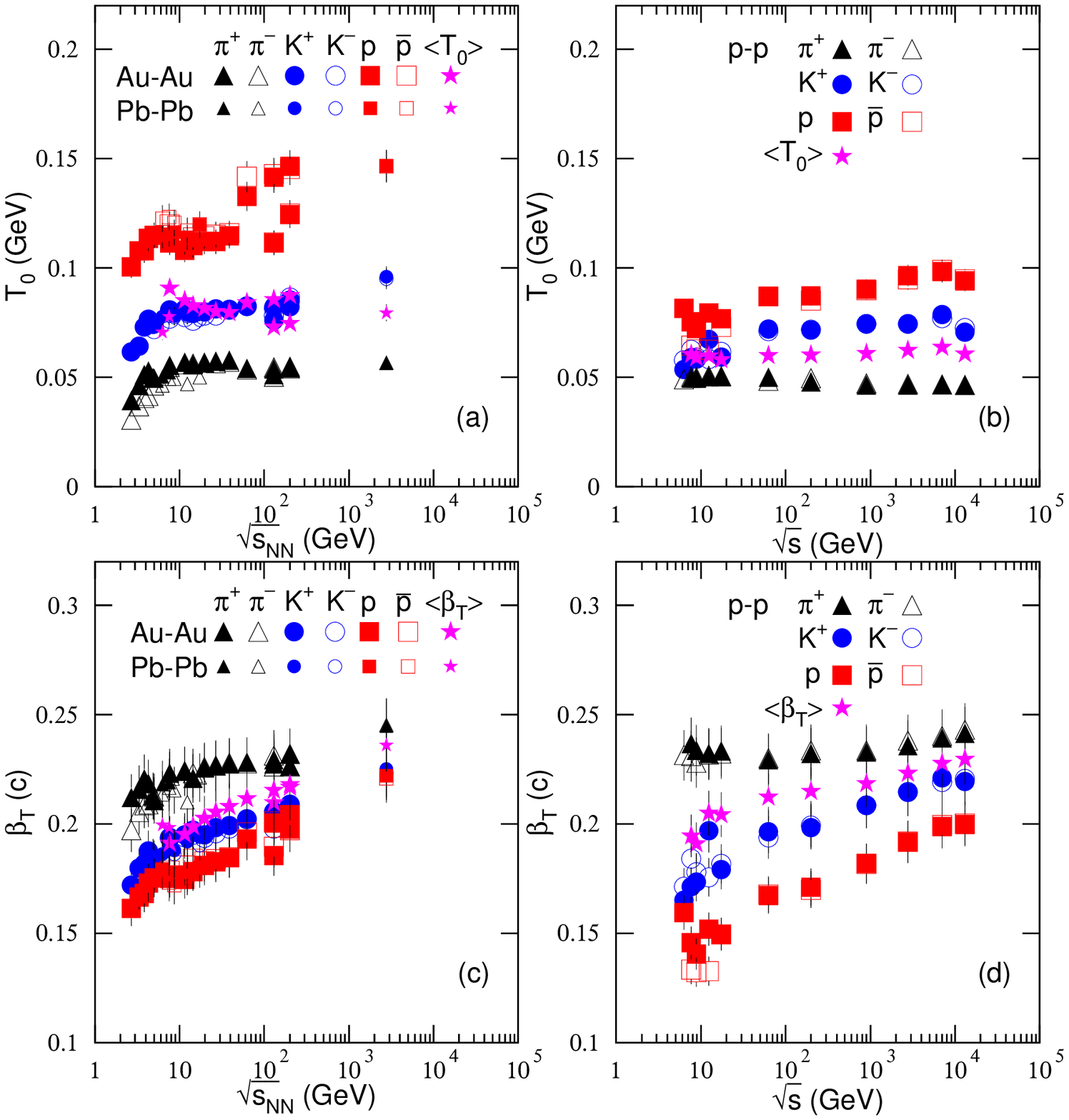}
\end{center}
{\small Fig. 5. Same as Fig. 4, but showing the excitation
functions of (a)(b) $T_0$ and $\langle T_0 \rangle$, as well as
(c)(d) $\beta_T$ and $\langle \beta_T\rangle$.}
\end{figure*}

\begin{figure*}[!htb]
\begin{center}
\includegraphics[width=16.0cm]{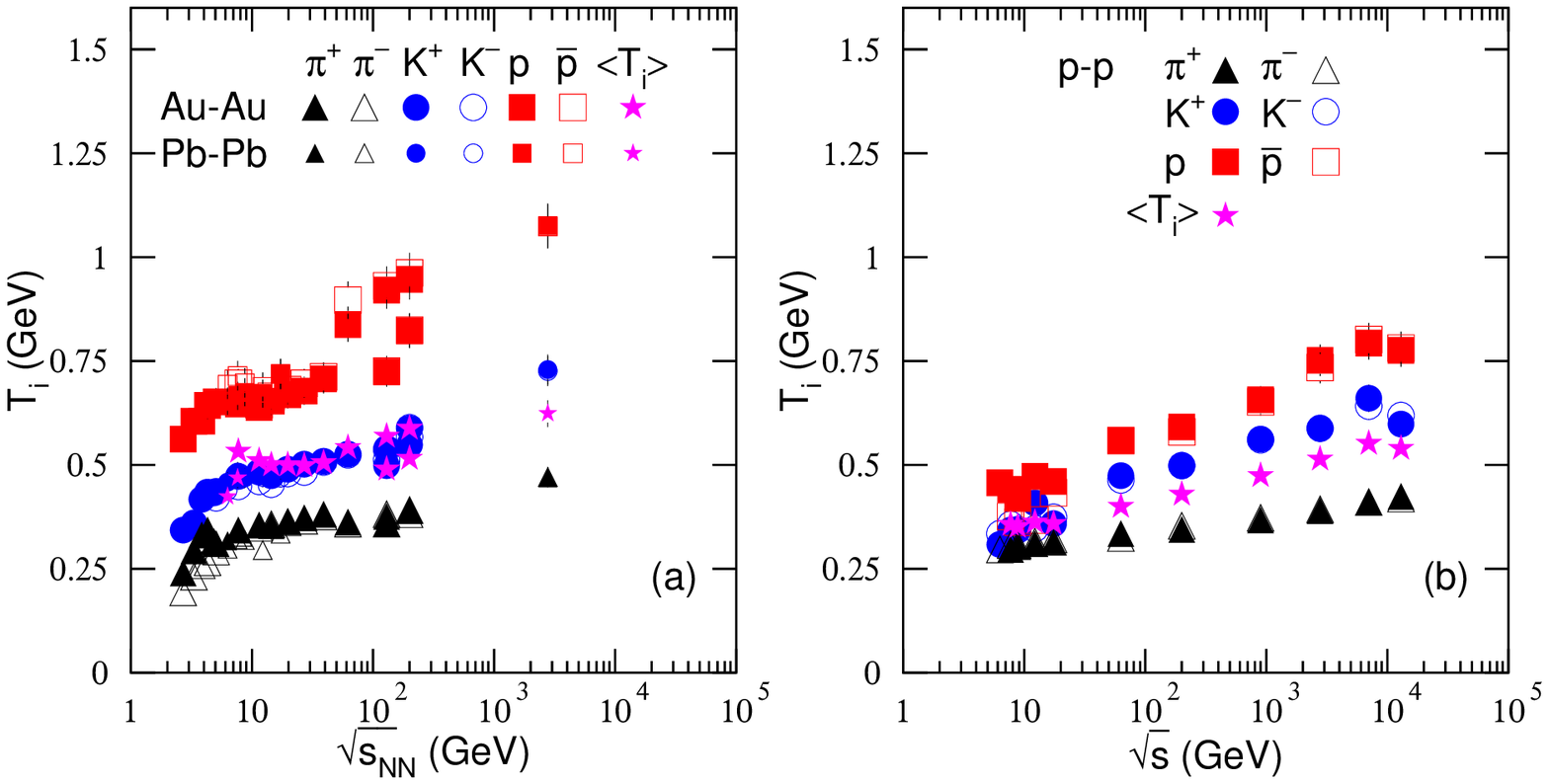}
\end{center}
{\small Fig. 6. Same as Fig. 4, but showing the excitation
functions of $T_i$ and $\langle T_i \rangle$.}
\end{figure*}

To better determine the kinetic freeze-out information, we now fit
simultaneously the spectra of $\pi^+$, $\pi^-$, $K^+$, $K^-$, $p$,
and $\bar p$ in different $p_T$ ranges using the same set of
parameters. In Figs. 1--3, the dot-dashed curves are the results
using the weighted average $\langle T\rangle$ which is energy
dependent, though in fact we may use other $T$ to obtain a little
better result in some cases. In the refit, the normalization
constant $N_0$ for different spectra is adjustable to fit suitable
$p_T$ range. In despite of mass dependent (two-)temperature in
non-simultaneous fit, the simultaneous fit is done using the same
and only set of $\langle T\rangle$ for the six species of
particles at each energy. That is to say that we use one component
function to fit the spectra of the six species of particles at
each energy. And the temperature is the same for each particle and
the normalization factor is adjusted. It seems that our treatment
is not a fair comparison of the non-simultaneous fit and
simultaneous fit, where the non-simultaneous fit uses the
two-component in some cases. However, the simultaneous fit seeks
for the same temperature which does not allow the two-component
for some particles due to other particles corresponding to one
component. In our opinion, our comparison is fair for the
non-simultaneous fit and simultaneous fit. One can see that the
same set of $\langle T\rangle$ can fit only a part of $p_T$ range
in some cases. The fit results using the same set of parameters
are not ideal. These results do not support the single scenario
for kinetic freeze-out. We are not inclined to fit simultaneously
the spectra of different particles. Conversely, we use different
$T$ for different spectra in this paper.

Figures 5(a) and 5(b) [5(c) and 5(d)] are similar to Figs. 4(a)
and 4(b) respectively, but they showing the excitation functions
of $T_0$ ($\beta_T$) and $\langle T_0\rangle$ ($\langle
\beta_T\rangle$). Figures 6(a) and 6(b) are also similar to Figs.
4(a) and 4(b) respectively, but they showing the excitation
functions of $T_i$ and $\langle T_i\rangle$. One can see that
$T_0$, $\langle T_0\rangle$, $\beta_T$, $\langle \beta_T\rangle$,
$T_i$, and $\langle T_i\rangle$ increase with the increase of
$\ln(\sqrt{s_{NN}})$ [$\ln(\sqrt{s})$] in general, and $T_0$ for
pion emission and $\langle T_0\rangle$ appears the trend of
saturation at the RHIC and LHC. Meanwhile, $T_0$, $\langle
T_0\rangle$, $T_i$, and $\langle T_i\rangle$ increase, and
$\beta_T$ and $\langle \beta_T\rangle$ decrease, with the increase
of particle mass.

It is regretful that some particles are absent in experimental
measurements at the energies blow 10 GeV. This renders that
$\langle T\rangle$, $\langle T_0\rangle$, $\langle
\beta_T\rangle$, and $\langle T_i\rangle$ are not available in the
energy range of several GeV. From the trends of available $T$,
$T_0$, $\beta_T$, and $T_i$, we may estimate that $\langle
T\rangle$, $\langle T_0\rangle$, $\langle \beta_T\rangle$, and
$\langle T_i\rangle$ increase (quickly) with the increase of
$\ln(\sqrt{s_{NN}})$ [$\ln(\sqrt{s})$] in the energy range of
several GeV. In particular, some excitation functions show little
peak around 10 GeV, which should be studied further in future.
Meanwhile, we hope to obtain more data in the energy range of
several GeV in future.

It should be noted that since empirical Eq. (18) is essential for
obtaining the energy dependence of $T_0$ and $\beta_T$, it seems
that one can obtain arbitrary results by choosing another
parametrization of $k_0$. In particular, it seems possibly to
choose a parametrization of $k_0$ so that $\pi^+$ ($\pi^-$), $K^+$
($K^-$), and $p$ ($\bar p$) could have similar $T_0$ and $\beta_T$
even though the combined temperature $T$ (in Fig. 4) for these
particles are quite different. This case corresponds to the single
scenario for kinetic freeze-out, which is not consistent with two
or multiple scenarios observed in other
studies~\cite{59a,59b,59c,59d}. We are inclined to use multiple
scenarios~\cite{59e} due to more accurate fit to the wider $p_T$
spectra. To coordinate the single and multiple scenarios, we may
regard the multiple scenarios as a refined situation of the single
scenario, which will be discussed later in detail.

The trends of excitation functions render that the collision
system undergoes different evolution processes. From several GeV
to about 10 GeV, the violent degree of collisions increases with
increasing the energy and matter density of the collision system.
The hadronic matter in the collision system stays at a state with
ever higher density and temperature. At about 10 GeV, the energy
and matter density of the collision system reaches to a high
value. The temperature is also high, which is needed to come into
notice. At above 10 GeV, the energy and matter density of the
collision system reaches to a higher value. The temperature is
also higher. However, because the phase transition from hadronic
matter to QGP had happened possibly, the temperature is limited,
which results in the levels of $T_0$ for pion emission and
$\langle T_0\rangle$ had stabilized.

Before summary and conclusions, we would like to point out that we
have used a new method to extract $T_0$, $\beta_T$, and $T_i$.
After fitting the $p_T$ ($m_T$) spectra by using the two-component
standard distribution Eqs. (6) or (7) in which the free parameters
are the effective temperatures $T_1$ and $T_2$ and the
contribution fraction $k$ of the first component, the derived
parameter, the effective temperature $T$, can be obtained from the
weighted average formula Eq. (8). Then, the derived parameters,
the kinetic freeze-out temperature $T_0$ and transverse flow
velocity $\beta_T$, can be obtained respectively from Eqs. (16)
and (17) which are related to the mean transverse momentum
$\langle p_T\rangle$. The derived parameter, the initial
temperature $T_i$, can be obtained from Eq. (29) which is related
to the root-mean-square $\sqrt{\langle p_T^2\rangle}$.

According to the analysis of the spectra of six hadron species
listed in Tables 1 and 2, one can see that pions, kaons, and
(anti)protons correspond to different temperatures of emission
source. This shows a mass-dependent multiple scenario for kinetic
freeze-out. Moreover, in $AA$ collisions at energies below the
LHC, charged pions can be redistributed between two sources, one
is hot and another is cold, while kaons and (anti)protons are
located in single (hot) sources (though with different
temperatures) in most cases. This is understandable. That charged
pions come from resonance decays contribute a relative large
fraction in low-$p_T$ region, which can be described by the first
component in Eq. (6) or (7). Some low-$p_T$ pions from
non-resonance decay can be also described by the first component.
As an ensemble, Eq. (1) describes the cold source with low $T$ for
all low-$p_T$ pions. However, the resonance decays for kaons and
(anti)protons are relatively small comparing to those for pions in
low-$p_T$ region, which are concealed in single source.

Naturally, if we regard $\langle T\rangle$, $\langle T_0\rangle$,
$\langle \beta_T\rangle$, and $\langle T_i\rangle$ as common
quantities corresponding to emissions of various hadron species,
we may use the mass-independent single scenario for kinetic
freeze-out and other system evolution stages such as chemical
freeze-out and initial state. It is contentious that the
mass-independent single scenario or mass-dependent multiple
scenario is right due to different physics thinkings. In our
opinion, the mass-independent single scenario is a very ideal
situation which is similar to the equilibrium state of mixture
gas. And the mass-dependent multiple scenario describes a refined
emission process which ``shows massive particles coming out of the
system earlier in time with smaller radial flow velocities, which
is hydrodynamic behavior"~\cite{60}. The temperatures discussed in
this paper reflect mainly the kinetic energies of various hadron
species, but do not have certainly the statistical sense.

We note that, in the mass-dependent multiple scenario for kinetic
freeze-out (Figs. 4 and 5), the obtained temperature for proton
emission is much larger than that for pion emission. This reflects
that protons coming out of the system is much earlier than pions
due to much larger mass of proton comparing to pion. This
phenomenon is a hydrodynamic behavior~\cite{60}, in which massive
particles are early leaved behind in the evolution process of
collision system. In other words, massive particles are not
emitted from the system on their owns initiative due to high
$T_0$, but are leaved behind under compulsion due to low $\beta_T$
and large $m_0$. In fact, some protons existed in projectile and
target nuclei appear in rapidity space as leading protons outside
the fireball. This issue also results in protons coming out of the
system to be earlier than pions.

Because $T_0$ and $\beta_T$ are model dependent, this paper is
different from Figs. 37 and 39 in ref.~\cite{30}, though this
paper is less model dependent and ref.~\cite{30} is much model
dependent. In ref.~\cite{30}, a flow velocity profile parameter
$n$ is used in the extraction of $T_0$ and $\beta_T$. The
parameter $n$ can be largely changed from 0 to 2 in $AA$
collisions and above $4$ in pp collisions, which is mutable and
debatable. The pion spectra in low-$p_T$ region ($<0.5$ GeV/$c$)
are excluded from the fit~\cite{30} due to resonance decay, which
overrates $T_0$ and $\beta_T$. Our work shows that $T_0$
($\beta_T$) in $pp$ collisions is slightly smaller than (almost
equal to) that in $AA$ collisions, which is in agreement with our
recent work~\cite{12} in which the intercept in the linear
relation of $T$ versus $m_0$ is regarded as $T_0$ and the slope in
the linear relation of $\langle p_T\rangle$ versus
$m_0\overline{\gamma}$ is regarded as $\beta_T$. This result is
understandable due to similar collective behavior as in $AA$
collisions appearing in $pp$ collisions~\cite{10}.

We would like to emphasize that this paper is a data-driven
reanalysis based on some physics considerations, but not a simple
fit to the data. From the data-driven reanalysis, the excitation
functions of some quantities such as the effective temperature $T$
and its weighted average $\langle T\rangle$, the kinetic
freeze-out temperature $T_0$ and its weighted average $\langle
T_0\rangle$, the transverse flow velocity $\beta_T$ and its
weighted average $\langle \beta_T\rangle$, as well as the initial
temperature $T_i$ and its weighted average $\langle T_i\rangle$
have been obtained. These excitation functions have appeared some
obvious laws with the increase of collision energy.

In the above discussions, to obtain $\overline\gamma$ then
$\beta_T$, we have used the MC method. Figs. 5(c) and 5(d) are a
direct result by the MC method. As a statistical model is
implemented for Figs. 1--3, we can also obtain the curves by Eq.
(19), which is in terms of the MC method. In fact, in the
calculation by the MC method, after the shuffled treatment due to
the randomicity by the Matlab code, we can obtain a lot of
``simulated data". Then, we may count them in different $p_T$
($m_T-m_0$) bins and obtain similar or the same results to the
curves in Figs. 1--3. In the case of the event numbers being not
too large, we shall observe fluctuations around the curves. As an
example, for $K^+$ spectra in Figs. 1(a)--1(f), the dotted curves
and crosses represent the MC results with high and low statistics
respectively. The two results from the analytical function and MC
method are confirmed each other.
\\

{\section{Summary and conclusion}}

We summarize here our main observations and conclusions.

(a) The transverse momentum or mass spectra of $\pi^+$, $\pi^-$,
$K^+$, $K^-$, $p$, and $\bar p$ at mid-$y$ or mid-$\eta$ produced
in central Au-Au (Pb-Pb) collisions over an energy range from 2.7
to 200 (6.3 to 2760) GeV have been analyzed in this work.
Meanwhile, the spectra in INEL $pp$ collisions over an energy
range from 6.3 to 13000 GeV have also been analyzed. In most
cases, the experimental data measured by the E866, E895, E802,
NA49, NA61/SHINE, STAR, PHENIX, ALICE, and CMS Collaborations are
approximately fitted by the (two-component) standard distribution
in which the temperature concept is the closest to the ideal gas
model.

(b) The effective temperature and its excitation function are
obtained from the transverse momentum or mass spectra of
identified particles produced in collisions at high energies. The
kinetic freeze-out temperature and transverse flow velocity and
their excitation functions are extracted from the formulas related
to the average transverse momentum, which is based on the
multisource thermal model. The initial temperature and its
excitation function are extracted from the formula related to the
root-mean-square transverse momentum, which is based on the color
string percolation model.

(c) With the increase of collision energy, the four derived
parameters and each average increase (quickly) from a few GeV to
about 10 GeV, then increases slowly after 10 GeV. In particular,
the kinetic freeze-out temperature for pion emission and its
average finally appear the trend of saturation at the RHIC and
LHC. Meanwhile, the three derived temperatures increases and the
derived transverse flow velocity decreases with the increase of
particle mass, which result in a mass-dependent multiple scenario
for kinetic freeze-out and other system evolution stages such as
chemical freeze-out and initial state.
\\
\\
{\bf Data Availability}

The data used to support the findings of this study are included
within the article and are cited at relevant places within the
text as references.
\\
\\
{\bf Ethical Approval}

The authors declare that they are in compliance with ethical
standards regarding the content of this paper.
\\
\\
{\bf Disclosure}

The funding agencies have no role in the design of the study; in
the collection, analysis, or interpretation of the data; in the
writing of the manuscript; or in the decision to publish the
results.
\\
\\
{\bf Conflicts of Interest}

The authors declare that there are no conflicts of interest
regarding the publication of this paper.
\\
\\
{\bf Acknowledgments}

This work was supported by the National Natural Science Foundation
of China under Grant Nos. 11575103 and 11947418, the Chinese
Government Scholarship (China Scholarship Council), the Scientific
and Technological Innovation Programs of Higher Education
Institutions in Shanxi (STIP) under Grant No. 201802017, the
Shanxi Provincial Natural Science Foundation under Grant No.
201901D111043, and the Fund for Shanxi ``1331 Project" Key
Subjects Construction.
\\

\end{multicols}
\end{document}